\documentclass[a4paper,11pt]{article}
\pdfoutput=1
\usepackage{jcappub}
\usepackage{ulem}
\usepackage{subfigure}

\newcommand{\by}{{\mathbf y}}
\newcommand{\bx}{{\mathbf x}}

\newcommand{\bn}{{\mathbf n}}

\newcommand{\bV}{{\mathbf V}}

\newcommand{\HH}{{\cal H}}

\newcommand{\de}{\delta}
\newcommand{\De}{\Delta}

\newcommand{\La}{\Lambda}
\newcommand{\la}{\lambda}
\newcommand{\Om}{\Omega}

\newcommand{\be}{\begin{equation}}
\newcommand{\ee}{\end{equation}}
\newcommand{\gsim}{\stackrel{>}{\sim}}

\newcommand{\bea}{\begin{eqnarray}}
\newcommand{\eea}{\end{eqnarray}}
\newcommand{\bean}{\begin{eqnarray*}}
\newcommand{\eean}{\end{eqnarray*}}
\newcommand{\dd}{\partial}

\newcommand*\bra[1]{\left(#1 \right)}
\newcommand*\sbra[1]{\left[#1 \right]}

\newcommand*\abs[1]{\left| #1 \right|}

\newcommand{\class}{{\sc class}}
\newcommand{\classgal}{{\sc class}gal}

\title{The \classgal{} code for Relativistic Cosmological Large Scale Structure}

\author[a]{Enea Di Dio, }
\author[a]{Francesco Montanari, }
\author[b,c,d]{Julien Lesgourgues, }
\author[a]{Ruth Durrer }

\affiliation[a]{D\'epartement de Physique Th\'eorique and Center for Astroparticle Physics, 
Universit\'e de Gen\`eve\\ 24 quai Ernest  Ansermet, 1211 Gen\`eve 4, Switzerland}
\affiliation[b]{Institut de Th\'eorie des Ph\'enom\`enes Physiques, \'Ecole Polytechnique F\'ed\'erale de Lausanne,\\ CH-1015, Lausanne, Switzerland}
\affiliation[c]{CERN, Theory Division,\\ CH-1211 Geneva 23, Switzerland}
\affiliation[d]{LAPTh (CNRS -Universit\'e de Savoie),\\ BP 110, F-74941 Annecy-le-Vieux Cedex, France}

\emailAdd{Enea.DiDio@unige.ch}
\emailAdd{Francesco.Montanari@unige.ch}
\emailAdd{Ruth.Durrer@unige.ch}
\emailAdd{Julien.Lesgourgues@cern.ch}

\abstract{
 We present accurate and efficient  computations of large scale structure observables, obtained with a modified version of the \class{} code which is made publicly available. This code includes all  relativistic corrections and computes both the power spectrum $C_\ell(z_1,z_2)$ and the corresponding correlation function $\xi(\theta,z_1,z_2)$ of the matter density and the galaxy number fluctuations in linear perturbation theory. For  Gaussian initial perturbations, these quantities contain the full information encoded in the large scale matter distribution at the level of linear perturbation theory. We illustrate the usefulness of our code for cosmological parameter estimation through a few simple examples.
}

\keywords{
Cosmology: Theory, Forecasts, Large Scale Structure
\vskip13pt plus8pt minus11pt
\noindent{\bfseries\large\sffamily{Preprints:}} CERN-PH-TH/2013-154, LAPTH-036/13
}
\arxivnumber{1307.1459}

\begin{document}
\maketitle
\flushbottom

\section{Introduction}
Since many years now, cosmology is a data driven science. This became especially evident with the discovery of 
 the apparent acceleration of the expansion of the universe,
which was found by observations and still remains very puzzling on the theoretical side. The most blatant success story of cosmology, however, remains the agreement between predictions and observations of the cosmic microwave background (CMB) anisotropies, see~\cite{Komatsu:2010fb,Larson:2010gs,Durrer:2008aa}, which is confirmed with the new Planck 
results~\cite{Ade:2013ktc,Ade:2013zuv}. 

We now want to profit also in an optimal way from actual and future galaxy catalogs which contain  information on
the large scale matter distribution, termed large scale structure (LSS). Contrary to the CMB which is  two dimensional, coming mainly from the  surface of last scattering, galaxy catalogs are three dimensional and therefore contain potentially more, richer information. On the other hand, galaxy formation is a complicated non-linear process, and it is not clear how much cosmological information about the underlying matter distribution and about gravitational clustering can be inferred from
the galaxy distribution. This is the problem of biasing which we do not address in this paper. Here we simply assume that on large enough scales, biasing is linear and local, an hypothesis which might turn out to be too simple~\cite{BeltranJimenez:2010bb}.

When observing galaxies, we measure their redshift $z$ and their angular position $-\bn = (\sin\theta\cos\phi,\sin\theta\sin\phi,\cos\theta)$. Note that $\bf{n}$ is the photon direction, so from the source to the observer. Hence we see a galaxy in direction $-\bf{n}$.  This observed three-dimensional data does not only contain information on the galaxy position, but also on the cosmic velocity field (redshift space distortions) and on perturbations of the geometry, e.g., lensing effects. Therefore, by making optimal use of galaxy
catalogs, we can learn not only about the large scale matter distribution but also about the velocity field and the geometry. Since Einstein's equations relate these quantities, this allows us  to test general relativity or more generally the $\La$CDM hypothesis, and to estimate cosmological parameters.

In this paper we present a new version of the Cosmic Linear Anisotropy Solving System (\class) code\footnote{\url{http://class-code.net}} \cite{Lesgourgues:2011re,Blas:2011rf}, incorporating several correction terms already presented in the theoretical works of Ref.~\cite{Yoo:2009au,Yoo:2010ni,Bonvin:2011bg,Challinor:2011bk,Bertacca:2012tp}. This code is called \classgal{} and is made publicly available on a dedicated website\footnote{\url{http://cosmology.unige.ch/tools/}}. The parts that concern the CMB have not been changed with respect to the main \class{} distribution. Several new features of \classgal{} will be merged with the main code in future \class{} versions.

In the next section, we describe the equations solved by \classgal{} and the initial input and final output. In section~\ref{s:pow}, we discuss the relevance of the different contributions to the observed power spectrum. In section~\ref{s:fish}, we present
some forecasts for parameter estimation with future catalogs,
 in order to illustrate the usefulness of our code. The topic of this section is worked out in more detail in an accompanying publication~\cite{Didio2}. In Section~\ref{s:con}, we conclude with an outlook to future possibilities using our code. The detailed description of the modifications of the \class{} code as well as some derivations are deferred to two appendices. 

\section{CLASS$\rm \bf gal$, a code for LSS}\label{s:code}

When we observe galaxies at a given redshift $z$ and direction $- \bn$, we cannot infer their position $\bx$. First of all, even in  an unperturbed
Friedman universe where $\bx= -r(z)\bn$, the radial comoving distance $r(z)$ depends on cosmological parameters. For small redshifts $z\ll 1$, we have simply $r(z)=zH_0^{-1}$, where $H_0$ denotes the present value of the Hubble parameter (the $H_0$ dependence can be removed by measuring distances in units of $h^{-1}$Mpc, where $H_0 = 100\,h\,\mathrm{km}\,\mathrm{s}^{-1}\mathrm{Mpc}^{-1}$). For  redshifts of order unity and larger, this approximation is no longer sufficient, 
 and one has to take into account the full time dependence of the Hubble parameter $H(z)$:
\bea\label{e:dist}
r(z) &=&  \int_0^z\frac{dz'}{H(z')} \,.
\eea
Normalizing  the scale factor $a$ to unity today, $a_0=1$, one has
\bea
H^2(z) &=& H_0^2\left(a^{-3}\Om_m+ a^{-2}\Om_K +\Om_\La\right) \, ,
\eea
where
\bea
\Om_m &=& \frac{8\pi G\rho_m(t_0)}{3H_0^2} \quad \begin{array}{l}\mbox{is the matter density parameter,} \end{array} \\
\Om_K &=& \frac{-K}{H_0^2} \quad \mbox{is the curvature parameter,}\\
\text{and} && \nonumber  \\
\Om_\La &=& \frac{\La}{3H_0^2} \quad \mbox{is the cosmological constant parameter.}
\eea
For a more complicated dark energy model, the expression involving $\Om_\La$ has to be modified correspondingly. 

Note that both $\bx$ and $r(z)$ are comoving distances. 
Hence, even in an unperturbed Friedmann universe, the three dimensional correlation function 
\bea
 \xi(|\by-\bx|,t) &=& 
\frac{\langle\rho(\bx,t)\rho(\by,t)\rangle}{ \bar\rho(t)^2} -1
\eea 
 depends on cosmological parameters, and so does its Fourier transform, the power spectrum $P(k,t)$.

But this is not all. The observed redshifts are perturbed by peculiar motions and by fluctuations of the geometry. The first is manifest e.g. in the well known redshift space 
distortions~\cite{Kaiser1987,Hamilton1992,Raccanelli:2010hk,Montanari:2012me}, while the latter is known e.g. through the effect of lensing on number counts~\cite{Lima:2010tq,Hezaveh:2010zk}. 

The gauge invariant expression for the perturbation of galaxy number counts, valid in a flat Friedmann universe, at first order in relativistic perturbation theory and ignoring bias, has been derived in Refs.~\cite{Yoo:2009au,Yoo:2010ni,Bonvin:2011bg,Challinor:2011bk}.  The result of~\cite{Bonvin:2011bg} for the
perturbation of number counts in direction $\bn$ and at redshift $z$ reads:
\bea
\De(\bn,z) &=& D_g +\Phi + \Psi + \frac{1}{\HH}
\left[\Phi'+\dd_r(\bV\cdot\bn)\right]  \nonumber \\  && 
+  \left(\frac{{\HH}'}{\HH^2}+\frac{2}{r_S\HH}\right)\left(\Psi+\bV\cdot\bn+ 
 \int_0^{r_S}\hspace{-3mm}dr(\Phi'+\Psi')\right) 
 \nonumber \\  &&   \label{Dez}
  +\frac{1}{r_S}\int_0^{r_S}\hspace{-3mm}dr \left[2 - \frac{r_S-r}{r}\Delta_\Om\right] (\Phi+\Psi) .
\eea
Here $\Psi$ and $\Phi$ are the Bardeen potentials or, equivalently, the temporal and spatial metric perturbations in the longitudinal gauge.
The gauge-invariant quantities $D_g$ and $\bV$ coincide respectively with the density fluctuations in the spatially flat  gauge and the peculiar velocity in the longitudinal gauge. $\HH(z) = H(z)a$ is the comoving Hubble parameter and $\Delta_\Om$ denotes the angular Laplacian. 
Primes denote derivatives with respect to conformal time $\tau$, to reflect the notations used in \class{} (while Ref.~\cite{Bonvin:2011bg} used $t$ for conformal time and dots for conformal time derivatives).
In the first two lines, all perturbations are evaluated at the coordinates $(\tau(z), -r_S(z) \bn)$ corresponding to the {\it unperturbed}
position of an object seen at a redshift $z$ in the direction $-\bn$. Inside the integral, metric perturbations are evaluated at conformal time $(\tau_0-r)$, where $\tau_0$ is the conformal age of the universe, and at comoving radius $r$. 

Eq.~(\ref{Dez}) is valid only for vanishing spatial curvature $K=0$. In the remainder of this paper, as well as in the present version of the code \class{gal} we restrict ourselves to this case.

Note that in longitudinal gauge, the second parenthesis of the second line of  Eq.~(\ref{Dez})  is simply $-\de z(1+z)^{-1}$, see Ref.~\cite{Bonvin:2011bg} for the general expression.
When writing this line, we neglect a possible evolution of the number of counts, i.e. we assume that $a^3\bar n_S$ is constant, where $\bar n_S$ denotes the background number density. Allowing for evolution, one should add in the first parenthesis  of the second line of  Eq.~(\ref{Dez}) 
the term~\cite{Challinor:2011bk} 
$$
\frac{d\ln(a^3\bar n_S) }{\HH d\tau_S}  = -(1+z)\frac{d}{dz}\ln\left(\frac{\bar n_S}{(1+z)^3}\right) \equiv f_{\rm evo}(z) \,.
$$
For $z\lesssim1.5$ evolution can be parametrized, e.g., using the Schechter luminosity function \cite{Zucca:2006,Peng:2010}. However, since $f_{\rm evo}(z)$ is very uncertain, we have set it to zero by default and simply allow the user to define his/her preferred evolution function $f_{\rm evo}(z)$, see Appendix~\ref{s:input} for more details.

Furthermore, as it stands, Eq.~(\ref{Dez}) gives the perturbation of the total number density. In practice, however, we cannot observe all galaxies, but only those with a flux which is larger than a certain limit, usually given in terms of a limiting magnitude $m_*$ related to a limiting flux $F_*$ by $m=-2.5\log_{10}F +\,$const. (the constant depends on the units in which we measure the flux). If the fluctuation of the source number density depends on luminosity, the number count at a fixed observed flux $F$ is given by 
\bea
\label{DezF}
\De(\bn,z,F)  &=&  \De(\bn,z) +\left.\frac{\dd\ln\bar n_S}{\dd \ln L_S}\right|_{\bar L_S}\times \frac{\de L_S}{\bar L_S} \nonumber \\
&=&  \De(\bn,z) +2\left.\frac{\dd\ln\bar n_S}{\dd \ln L_S}\right|_{\bar L_S} \times 
\left[\left(\frac{1}{r_S\HH}-1\right)\left(\Psi +\int_0^{r_S}dr(\Phi'+\Psi') +\bV\cdot \bn\right)
\right.   \nonumber\\
&& + \left.
 \frac{1}{2r_S}\int_0^{r_S}\hspace{-3.6mm}dr \left[2-\frac{r_S-r}{r}\Delta_\Om\right] (\Phi+\Psi) \,   -\Phi\right]  .
\eea
up to some local monopole and dipole terms that we neglect for consistency.
Here $\bar L_S$ is the background luminosity corresponding to a flux $F$.
In the second equality, we made use of the fact that the fractional fluctuation in the luminosity at fixed flux is given by twice the fractional fluctuation in the luminosity distance which is computed e.g. in~\cite{Bonvin:2005ps}. In Appendix~\ref{Lum_fluct_Appendix} we show that the result of Ref.~\cite{Bonvin:2005ps} is equivalent to the big bracket of Eq.~(\ref{DezF}).

Denoting the sources in direction $- \bn$ at redshift $z$ with magnitude $m<m_*$, i.e., flux $F>F_*$ by 
$N(\bn,z,m<m_*)dzd\Om_0$ , and its fractional perturbation by $D_g(L>\bar L_*)$, we then obtain for its fluctuation, see~\cite{Challinor:2011bk}
\bea
\De^{(N)}(\bn,z,m_*) &=& D_g(L>\bar L_*) +(1+5s)\Phi + \Psi + \frac{1}{\HH}
\left[\Phi'+\dd_r(\bV\cdot\bn)\right] +  \nonumber \\  &&
 \left(\frac{{\HH}'}{\HH^2}+\frac{2-5s}{r_S\HH} +5s-f^N_{\rm evo}\right)\left(\Psi+\bV\cdot\bn+ 
 \int_0^{r_S}\hspace{-0.3mm}dr(\Phi'+\Psi')\right) 
   \nonumber \\  &&  \label{DezNF}
+\frac{2-5s}{2r_S}\int_0^{r_S}\hspace{-0.3mm}dr \left[2-\frac{r_S-r}{r}\Delta_\Om\right] (\Phi+\Psi) \,.
\eea
with
$$
f_{\rm evo}^N = \frac{\partial\ln\bra{a^3 \bar N(z,L>\bar L_*)}}{\HH \partial \tau_S} \,.
$$
Here we have introduced the dependence of the number density on the luminosity via the  logarithmic derivative,
\be
s(z,m_*) \equiv \frac{\dd\log_{10} \bar N(z,m<m_*)}{\dd m_*} = \frac{\bar n_S(z,\bar L_*)}{2.5\bar N(z,L_S>\bar L_*)} \,,
\ee
where
\bea
\bar N(z,L_S>\bar L_*) \equiv \frac{\ln10}{2.5} \int_{-\infty}^{m_*} \bar n_S(z,m) dm  
= \int_{F_*}^{\infty} \bar n_S(z,\ln F) d\ln F \,.
\eea
Using this definition and the fact that at fixed $z$ partial derivatives w.r.t. $L$ are the same as those w.r.t $F$, one deduces:
$$\left. \frac{\partial \ln \bar N(z,L_S>\bar L_*) }{\partial\ln L_S} \right|_{\bar L_*} = -\frac{5}{2} s(z,m_*) \,,$$
which has been used to pass from Eqs.~(\ref{Dez},\ref{DezF}) to (\ref{DezNF}).
If the number density is independent of luminosity, $s$ vanishes, and if we can neglect evolution, $f_{\rm evo}=0$. Then Eq.~(\ref{DezNF}) reduces to Eq.~(\ref{Dez}). In the present version of the \classgal{} code, the user can introduce a  constant value for $s(z,m_*)$, depending on the limiting magnitude of the catalog she/he wants to analyze. The default value is $s=0$.

Note that the dominant contribution to the redshift space distortions, the last term on the first line of~(\ref{Dez}), does not dependent on luminosity nor evolution, while the lensing term (third line of~(\ref{Dez})) is affected by the luminosity dependence. This is the so called magnification bias.

In the code, instead of $D_g$, it is more convenient to use another gauge-invariant quantity $D$ that coincides with the density fluctuations in comoving gauge. For non-relativistic matter it is related to $D_g$ through~\cite{Durrer:2008aa}
\be \label{e:DgD}
D_g=D-3\left( \frac{\cal H}{k} V + \Phi \right)~.
\ee
Without mentioning it, $D_g$, $D$ and $V$ are always the corresponding quantities for matter (i.e. baryons, cold dark matter and possibly non-relativistic neutrinos) for which we set $w_m =p_m/\rho_m =0$.
We never use the perturbed Einstein equations in these expressions, so that it is easy to change the code when one wants to consider a different dark energy model which may itself have perturbations, or models which modify gravity. 

For the case of $\La$CDM, the Einstein equations in the matter and $\La$ dominated era are simply~\cite{Durrer:2008aa}
\be
4 \pi G a^2 \rho_\mathrm{tot} D_\mathrm{tot} = \frac{3}{2}\HH^2\frac{\Om_m}{\Om_m + \Om_\La a^3}D= -k^2\Phi~,  \qquad \mbox{and }~~ \Phi = \Psi \,.
\ee
The matter density perturbation $D$  in comoving gauge is related to the total density fluctuation it via
$ \left.\de\rho_m/\rho_\mathrm{tot}\right|_{\rm com} \equiv D_\mathrm{tot}  =\frac{\Om_m}{\Om_m + \Om_\La a^3}D$.

We can expand Eq.~(\ref{Dez}) or (\ref{DezNF}) in spherical harmonics with redshift dependent amplitudes,
\be
\De(\bn,z) =\sum_{\ell m}a_{\ell m}(z)Y_{\ell m}(\bn), ~\mbox{with }~  
a_{\ell m}(z) = \int d\Om_{\bn}Y_{\ell m}^*(\bn)\De(\bn,z).
\ee
The star indicates complex conjugation. Denoting the angular power spectrum by
\be
 C_\ell (z_1,z_2) = \langle a_{\ell m}(z_1) a^*_{\ell m}(z_2) \rangle 
\ee 
a short calculation gives~\cite{Bonvin:2011bg}
\be
 C_\ell (z_1,z_2) = 4 \pi \int \frac{dk}{k} {\cal P}(k) \De_\ell\left(z_1,k\right)\De_\ell\left(z_2,k\right) \label{Cl1}
 \ee
 where ${\cal P}(k)$ is the primordial power spectrum, and
 \bea
&&\hspace*{-4mm}\De_\ell(z,k) = 
 j_\ell(kr(z)) \left[ b D(\tau(z),k) + \left(\frac{{\cal H}'}{{\cal H}^2}+\frac{2-5s}{r(z) {\cal H}} +5s - f^N_{\rm evo} +1 \right)\!\Psi(\tau(z),k) \nonumber \right. \\
&&\left. \hspace{1.5cm}+~\left(  -2 +5s \right) \Phi(\tau(z),k) + {\cal H}^{-1} {\Phi}'(\tau(z),k) \right]  \nonumber \\
&&+ \left[ \frac{d j_\ell}{dx}(kr(z)) \!\left( \frac{{\cal H}'}{{\cal H}^2} + \frac{2 -5s}{r(z) {\cal H}} +5s\!- \!f^N_{\rm evo} \right)+ \frac{d^2 j_\ell}{dx^2}(kr(z)) \frac{k}{\cal H} \right. \nonumber \\
&&\left. \hspace{.5cm}+(f^N_{\rm evo}-3) j_\ell(kr(z))\frac{\HH}{k}\right]  V(\tau(z),k) \nonumber \\
&&+ \int_0^{r(z)} \hspace*{-0.4cm} dr \, j_\ell(kr) \left[ \left( {\Phi}(\tau,k) + {\Psi}(\tau,k) \right) \left( \frac{2-5s}{2}\right)\left(\!\ell(\ell+1) \frac{r(z)-r}{r(z) r} + \frac{2}{r(z)} \right) \right. \nonumber \\
&&\hspace{25mm}
+ \left. \left(\! {\Phi}'(\tau,k)+{\Psi}'(\tau,k) \!\right) \left(\!\frac{{\cal H}'}{{\cal H}^2} + \frac{2-5s}{r(z) {\cal H}} +5s -f^N_{\rm evo}\!\right)_{r(z)}
\right]\,.
\label{Fl}
 \eea
Here the $j_\ell(x)$'s are the spherical Bessel functions, and all perturbations are the real transfer functions relating the corresponding variables to the power spectrum (we give more details about the definition of the primordial spectrum and of all the transfer functions in Appendix~\ref{conventions}).
The times $\tau(z) = \tau_0-r(z)$ and $\tau=\tau_0-r$ are conformal times of perturbations which we see  at comoving distances $r(z)$ and $r$ respectively.
Adding a linear scale-independent bias $b$ between the comoving matter and galaxy density, taking into account galaxy evolution and using Eq.~(\ref{e:DgD}), we replace $D_g$ by $b D + (f^N_{\rm evo}-3)\HH V/k - 3 {\Phi}$ \cite{Challinor:2011bk}.

The choice of adding galaxy bias to the density in the comoving gauge $D$ is justified by the assumption that both galaxies and dark matter follow the same velocity field as they experience the same gravitational acceleration, and the linear bias prescription is valid in their rest frame. In the numerical calculations of the present paper we assume for simplicity $b=1$, for a detailed discussion of galaxy bias in General Relativity see \cite{Baldauf:2011bh}.
 
The first term in the first line of Eq.~(\ref{Fl}) is the usual density term. The third line collects all redshift space distortions terms and Doppler terms. The usual redshift space distortion derived by Kaiser~\cite{Kaiser1987} is  the term proportional to $[d^2j_\ell/dx^2] k/\HH$, but  sometimes, the subdominant term proportional to $2[d j_\ell/dx] /(r\HH)$ is also considered as part of the redshift space distortion.
The first term on the fourth line is the lensing term. Note that this term is parametrically of the same order as the density and redshift space distortion terms, and therefore can become important in certain situations. The other terms are sometimes called ``relativistic corrections''. This name is somewhat misleading, since redshift space distortions are also (special) relativistic contributions, and of course the lensing term is also relativistic. We shall therefore simply call them ``gravitational potential terms''. They contain an integrated Sachs Wolfe effect (term on the last line), and several contributions from the potentials $\Phi$ and $\Psi$ at redshift $z$. 
  
In this equation, we have neglected terms evaluated at $z=0$,  at the observer position, which contribute only to the monopole. We have also neglected the term induced by the observer velocity, which contributes only to the dipole. These terms cannot be calculated reliably within linear perturbation theory. Nevertheless, $C_0(z_1,z_2)$ and $C_1(z_1,z_2)$  contain some interesting information on clustering, to which these local contributions would only add an uninteresting $z$-independent constant, see~\cite{Didio2}.
 
Equations (\ref{Cl1}, \ref{Fl}) are valid for a pair of infinitely thin shells located at redshift $z_1$ and $z_2$.
For realistic  redshift bins of finite thickness  described by a set of window functions $W_i(z)$, e.g. a Gaussian centered at some redshift $z_i$, and a given number density of galaxies per redshift interval $dN/dz$ (the integral of the product $W_i(z) dN/dz$ being normalised to unity), one can substitute $\De_\ell(z,k)$ with the integral 
\be
\De_\ell^i(k) = \int dz \frac{dN}{dz} W_i(z) \De_\ell(z,k)~, \label{delta}
\ee
and define the power spectrum
\be
 C_\ell^{ij}  = 4 \pi \int \frac{dk}{k} {\cal P}(k)  \De_\ell^i (k)\De_\ell^j(k)~. \label{Cl2}
 \ee
For $i=j$, this quantity represents the auto-correlation power spectrum of relativistic density fluctuations observed in the shell around $z_i$. For $i\neq j$, it represents the cross-correlation power spectrum between two shells. When the window functions are Dirac distributions, we recover the expression  $C_\ell (z_i,z_j)$.

Our code \classgal{} calculates $C_\ell^{ij}$ for a given cosmological model and for a set of window functions (that can be chosen to be Gaussian, Dirac or top-hat distributions). We continue to use the notation $C_\ell (z_i,z_j)$ in the case of Dirac window functions.  We easily obtain angular correlation functions by summing up the $C_\ell$'s,
\bea
\xi(\theta,z_i,z_j)  &\equiv& \langle \De(\bn,z_i) \De(\bn',z_j)\rangle \nonumber  \\
&=& ~\frac{1}{4\pi}\sum_{\ell=0}^{\ell_{\max}} (2\ell+1)C_\ell (z_i,z_j)P_\ell(\cos\theta) W_{\ell} \,, \label{e:corfun}
\eea
where $W_{\ell}=\exp\sbra{-\ell(\ell+1)/\ell_s^2}$ is a Gaussian smoothing
introduced for convenience, which smoothes out sufficiently high multipoles
to avoid unphysical small scales oscillations in $\xi(\theta)$ which are an artifact coming from the finiteness of $\ell_{\max}$. For $\ell_{\max}\sim 1000$ it is sufficient to choose $\ell_s\sim 600$.
$P_\ell(\mu)$ is the Legendre polynomial of degree $\ell$ and $\theta$ is the angle between $\bn$ and $\bn'$, i.e., $\cos\theta =\bn\cdot\bn'$.
For the case of thick shells,
\be 
\xi(\theta)^{ij} =  \frac{1}{4\pi}\sum_{\ell=0}^{\ell_{\max}} (2\ell+1)C_\ell^{ij}  P_\ell(\cos\theta) W_{\ell}\,.
\label{eq:xi_theta}
\ee

\section{Power spectra and correlation functions}\label{s:pow}

In this section we show how \classgal{} can be used to estimate cosmological parameters. Of course, a truly measured correlation function will always have finite redshift and angular resolution,
that can be described respectively by the shape and width of the window function, and by an appropriate instrumental noise function growing exponentially  above a given $\ell$ or below a given angle.
Furthermore, since we measure density fluctuations with a discrete tracer, namely galaxies, we must add Poisson noise to the error budget.

We first study the sensitivity of the different terms in the power spectrum to the redshift resolution. 
Throughout this section we set $s=f_{\rm evo}=0$ in Eq.~(\ref{Fl}). We use a galaxy density distribution $dN/dz$ inspired from the characteristics of a survey like DES\footnote{\url{www.darkenergysurvey.org}},
\be
\frac{dN}{dz} \propto \bra{\frac{z}{0.55}}^2 \exp\sbra{-\bra{\frac{z}{0.55}}^2} ~.
\ee
This assumption is convenient for the purpose of comparing our results with those of Ref.~\cite{Asorey+12}. As explained in the second section of Appendix~\ref{app:A}, \classgal{} allows the user to pass any selection function $dN/dz$ analytically or in the form of tabulated values.
For each redshift bin associated with a window function $W_i(z)$, the product $W_i(z) dN/dz$ is normalized to unity.

\begin{figure}[tbp]
\begin{center}
\subfigure[]{
  \includegraphics[width=.47\columnwidth]{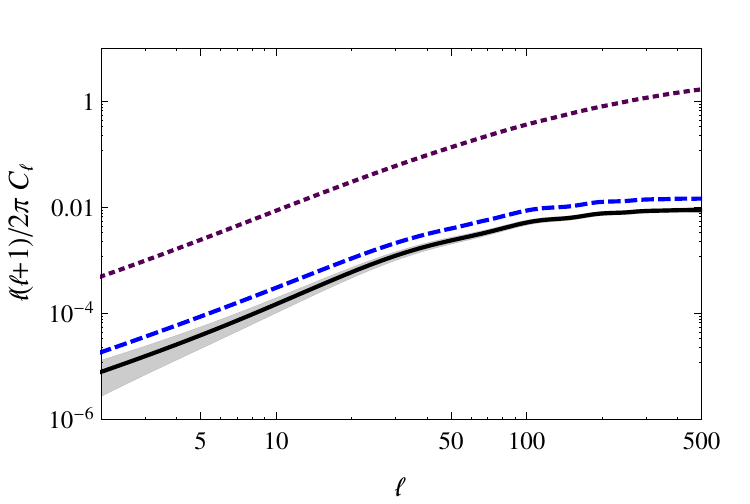}
  \label{fig:Cl_Windows}
}
\subfigure[]{
  \includegraphics[width=.47\columnwidth]{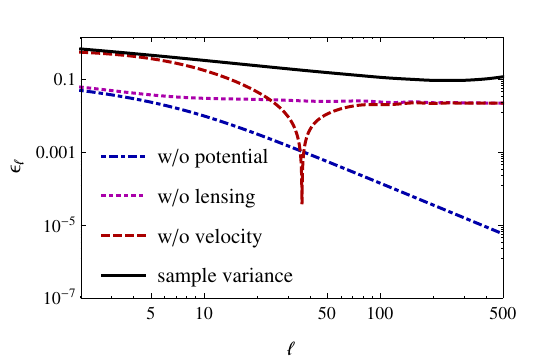}
  \label{fig:Cl_errors}
}
\end{center}
\caption{\emph{Left panel:} auto-correlation power spectrum, defined in Eq. (\ref{Cl2}), for a redshift bin centered around $z=0.55$ and three different choices of window function: a Gaussian with half-width $\De z=0.1$ (solid, black), a tophat of
half-width $\De z = 0.1$ (dashed, blue), and a Dirac delta (dotted, purple). For the Gaussian window function we also indicate the error $\sigma_{C_\ell}$ as a shaded band.\newline
\emph{Right panel:} errors in Eq. (\ref{Cl2}) when neglecting gradually gravitational potential terms
(dot-dashed, blue), lensing (dotted, purple) and velocity terms (Doppler and $z$-space distortions, dashed, red).
The r.m.s. variance $\sigma_{C_{\ell}}/C_{\ell}^{\rm full}$ is also shown (solid, black).
Results are computed for a Gaussian bin centered around $\bar z=0.55$ and with half-width 
$\De z=0.1$.
}
\end{figure}

Fig.~\ref{fig:Cl_Windows} shows the auto-correlation power spectrum of a given redshift bin $i$, defined in Eq.~(\ref{Cl2}), for different types
of window functions. In the case of a Gaussian window (solid curve) we plot also the r.m.s. variance given by
\be
\sigma_{C_{\ell}} = \sqrt\frac{2}{(2\ell+1)f_{\rm sky}} \bra{ C_{\ell}^{ii} + \frac{1}{n_{i}} } \;,
\ee
where $n_i$ is the number of galaxies per steradian inside the bin, and we 
assume full sky coverage, $f_{\rm sky}=1$. 
We assumed the same value of $n_i$ as in the table presented in Ref.~\cite{Asorey+12}, reflecting the characteristics of DES (see also footnote~\ref{footnoteSN}).

The Dirac delta case corresponds to a purely  theoretical quantity.
This signal is reduced significantly after the integration with a tophat or Gaussian
window function. The tophat can be used if galaxy redshifts are known with spectroscopic
precision. If we use a redshift resolution of $\De z =0.001$, the signal is reduced roughly by a factor of 2 with respect to the Dirac delta window function.
If only photometric redshift are available, then because of the relatively low
$z$-resolution, the sharp tophat has to be replaced by a Gaussian.

Fig.~\ref{fig:Cl_errors} shows the errors made when neglecting one or several terms in Eq.~(\ref{Fl}), for a Gaussian bin centered around $\bar z=0.55$ and with half-width $\De z=0.1$.
We define
\be
\epsilon_{\ell} = \abs{1-C_{\ell}^{\rm partial}/C_{\ell}^{\rm full}} ~,
\ee
where $C_{\ell}^{\rm full}$ is calculated with all the terms in Eq.~(\ref{Fl}),
while in $C_{\ell}^{\rm partial}$ we gradually neglect the gravitational potential terms (G1-G5 in Eq.~(\ref{eq:delta_terms})), the lensing term (`Len' in Eq.~(\ref{eq:delta_terms})), and all velocity terms (redshift-space distortions and Doppler, `Red' and `Dop' in Eq.~(\ref{eq:delta_terms})).
Similar comparisons have been presented in~\cite{Bonvin:2011bg,Challinor:2011bk} with a lower $\ell_\mathrm{max}$ and different choices of redshift and window functions.
Note that when we neglect a contribution,
we remove the corresponding auto-correlation as well as the correlations with other terms.
Hence the difference $\epsilon_{\ell}$ can become negative if it is dominated by a negative correlation term.
Spikes toward $-\infty$ (in log scale) arise when the difference changes sign.

For comparison, we also show the r.m.s. sample variance
$\sigma_{C_{\ell}}/C_{\ell}^{\rm full}$ (note that this quantity can in
principle be larger than 1). As expected, this variance is more important for lower multipoles.
After $\ell\approx300$ it increases again because shot-noise starts dominating
the error.

As expected, gravitational potential terms (dot-dashed) are always subdominant and most important on large scales.
The lensing term (dotted) is nearly scale-independent.
For the chosen configuration, namely a Gaussian window function centered at $\bar z=0.55$ and with half-width $\De z = 0.1$,
redshift-space distortions (dashed) represent the most important corrections to the plain density term,
and neglecting these distortions introduces an error of the same order as the sample
variance (solid) on large scales. At $\ell\gsim 38$, the difference $\epsilon_{\ell}$ from redshift-space distortions  changes sign and then become comparable to the lensing corrections.
We stress, however, that we are not seeing here the non-linear Fingers-of-God effect,
which is important on much smaller scales.

In a companion paper~\cite{Didio2}, we also consider situations where the redshift space distortion or the lensing term become significantly larger than the noise. There we see that wide window functions significantly reduce redshift space distortions but enhance the lensing term. Using single tracers, we have not yet found a configuration which is such that the potential terms raise above the noise.
However, since they are largest on large scales, they are significantly affected by cosmic variance. 
A multi-tracer analysis may be considered to reduce cosmic variance \cite{McDonald:2008sh}.
Recently, it has also been shown that asymmetries in the correlation function between bright and faint galaxies are useful to enhance the relativistic terms \cite{Bonvin2013}.

\begin{figure}[tbp]
\begin{center}
\subfigure[]{
  \includegraphics[width=.47\columnwidth]{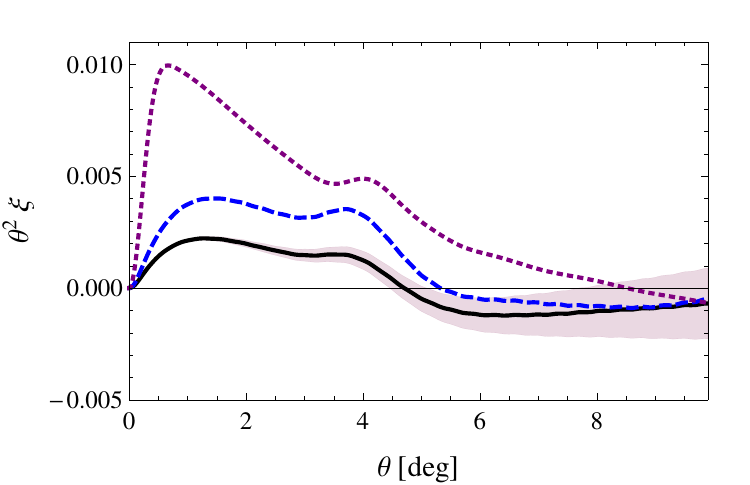}
  \label{fig:xi_Windows}
}
\subfigure[]{
  \includegraphics[width=.47\columnwidth]{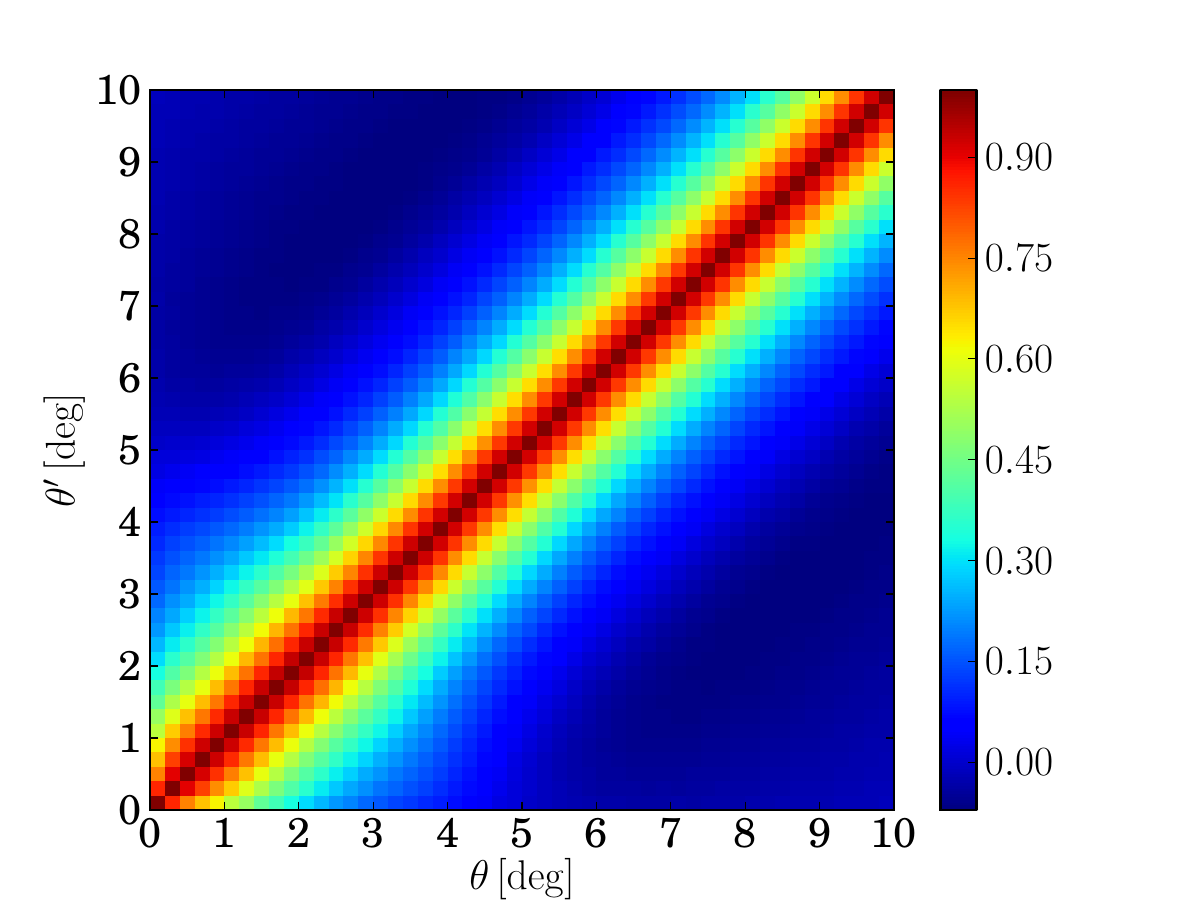}
  \label{fig:covThThp}
}
\subfigure[]{
  \includegraphics[width=.6\columnwidth]{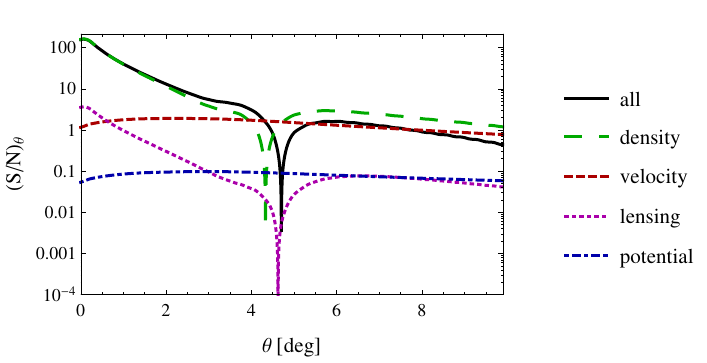}
  \label{fig:xi_errors}
}
\end{center}
\caption{\emph{Upper left:} the angular auto-correlation function, Eq. (\ref{eq:xi_theta}), for a bin centered at
$\bar{z}=0.55$ and for a Gaussian window function with $\De z=0.1$ (solid, black), a
tophat of half-width $\De z =0.1$ (dashed, blue), and a Dirac delta (dotted, violet). For the Dirac delta case the correlation function has been divided by 10 for easier comparison.
For the Gaussian window function we also indicate the error $\sigma_{\xi_{\theta}}$ as a shaded band. The bumps visible near four degrees correspond to the baryon acoustic feature.
\emph{Upper right:} the reduced covariance matrix of the angular correlation function (given by Eq. (\ref{e:covTh}) rescaled by $\sigma_{\xi_{\theta}}\sigma_{\xi_{\theta'}}$).
\emph{Bottom:} the signal of the different contributions relative to the r.m.s. variance of the full $\xi_{\theta}$. Results are computed for a Gaussian bin centered at $\bar z=0.55$ and with half-width $\De z=0.1$.
}
\end{figure}

Using Eq.~(\ref{eq:xi_theta}), we can also compute the angular correlation
function. Fig.~\ref{fig:xi_Windows} shows the auto-correlation in a redshift bin with either Gaussian, tophat or
Dirac window function. In the Dirac case, the result has been divided
by 10 for easier comparison using linear scales.
In the Gaussian case, we also plot the r.m.s. sample variance.

Observational data on the angular correlation function must be interpreted with care since points at different angles are correlated. On the theoretical level, their non-diagonal covariance matrix is given by~\cite{Crocce:2011}
\be
\label{e:covTh}
{\rm Cov}_{\theta\theta'} = \frac{2}{f_{\rm sky}} \sum_{\ell\geq0} \frac{2\ell+1}{(4\pi)^2} P_{\ell}(\cos\theta) P_{\ell}(\cos\theta') \bra{C_{\ell}+\frac{1}{n}}^2 ~,
\ee
and the error at a given angular scale can be estimated as
\be \label{eq:xi_var}
\sigma_{\xi_{\theta}} = \bra{ {\rm Cov}_{\theta\theta} }^{1/2} \,.
\ee

In Fig.~\ref{fig:covThThp} we show the
non-diagonal structure of the reduced covariance matrix defined as
${\rm Cov}_{\theta\theta'}/\sigma_{\xi_{\theta}}\sigma_{\xi_{\theta'}}$,
for a Gaussian window with half-width $\De z = 0.1$ centered at $\bar z=0.55$. 
Along the diagonal, the reduced covariance matrix is equal to one by construction.
Notice the small anti-correlations for $|\theta-\theta' | \gsim 6^o$ in dark blue.

Like for the power spectrum, we illustrate the impact of the different terms which appear in~Eq.~(\ref{Fl}) on the total correlation function.
Results are computed for a Gaussian bin centered around $\bar z=0.55$ and with half-width 
$\De z=0.1$.
Fig.~\ref{fig:xi_errors} shows the signal of different effects compared to the r.m.s. variance of the full correlation function:
\be \label{eq:xi_SN}
\bra{\frac{S}{N}}_\theta = \frac{\xi(\theta)^{\rm partial}}{\sigma_{\xi_{\theta}}} \;,
\ee
where the nominator takes into account density, velocity terms (Doppler and $z$-space distortions), lensing and potential effects as well as their cross-correlations with previous terms, respectively.
Therefore, e.g., the lensing curve considers not only lensing-lensing correlations but also cross-correlations of lensing with velocity and density terms.
The curve corresponding to the total, observable, $\xi(\theta)$ is also plotted.
The denominator of Eq.~(\ref{eq:xi_SN}) is given by Eq.~(\ref{eq:xi_var}), which is computed considering the contribution of all the terms.
Density and lensing curves as well as the total correlation function exhibit a spike to $-\infty$ (in log scale) at the angle for which $\xi(\theta)$ crosses zero. Note that the zero of the full correlation function moves from about $4.2^o$ to $4.6^o$ due to the presence of redshift space distortions. One also sees clearly that on large scales, $\theta>4.5^o$ density and redshift space distortions are anti-correlated while on small scales they are correlated.

The $(S/N)_\theta$ curves show whether an effect gives a contribution larger than sample variance at a given scale, hence $(S/N)_\theta>1$ suggests that in principle it the corresponding contribution is observable if it can be isolated from the other terms.
The main contribution comes from density and redshift space distortions, consistent with Fig.~\ref{fig:Cl_errors}.
Lensing is mainly important at small scales, where linear perturbation theory which is adopted here is not longer sufficient since angles $\theta\lesssim 1^\circ$ correspond to comoving separations $r\lesssim 25\, {\rm Mpc}/h$.
Potential terms have a signal-to-noise which is  nearly scale-independent. Therefore they are more relevant at large separations where other effects decay. Nevertheless, due to cosmic variance $(S/N)_\theta$ of the potential terms never raises towards 1.

The code {\sc camb}sources\footnote{\url{http://camb.info/sources/}} \cite{Challinor:2011bk} allows to calculate
$C_\ell(z_1,z_2)$ for not too narrow smooth window functions and at not too large  values of $\ell$~\footnote{The version of {\sc camb}sources available at the time of this writing (October 2013) leads either to instabilities or to prohibitive memory requirements in the limit of thin Gaussian shells and/or large $\ell$.}. 
Whenever possible, we have checked that our results for the $C_{\ell}$'s agree with the output of {\sc camb}sources,
and that those for $\xi(\theta)$ agree with \cite{Montanari:2012me}.
Our code \classgal{} has been optimized also for non-Gaussian window functions, narrower redshift bins and higher $\ell$. It includes several extra options with respect to {\sc camb}sources, that are described in section \ref{s:code} of Appendix~\ref{app:A}.

\section{Example}\label{s:fish}

As an example, we now determine the accuracy of $\Om_m$ obtained from a galaxy survey, keeping all other parameters fixed, but varying the number of redshift bins $N_\text{Bins}$ in which we split the data. 
The Fisher matrix  is defined as
\be
\label{eq:fisher}
F_{\alpha \beta} = \sum_{\ell} \frac{\partial C_\ell^{ij} }{\partial \lambda_\alpha} \frac{\partial C_\ell^{pq}}{\partial \lambda_\beta} \text{Cov}^{-1}_{\ell, (ij), (pq)},
\ee
where the covariance matrix of a given survey with sky coverage $f_\text{sky}$ reads
\be
\text{Cov}_{[\ell,\ell'] [(ij), (pq)]}=\de_{\ell,\ell'}\frac{C_\ell^{\text{obs},i p} C_\ell^{\text{obs},jq} + C_\ell^{\text{obs},i q} C_\ell^{\text{obs},jp}}{f_\text{sky} \left( 2 \ell + 1 \right) },
\ee
and the spectra $C_\ell^{\text{obs} ,ij}$ include a shot-noise contribution related to $n(i)$, the number of galaxies per steradian in the $i$-th redshift bin,
\be
C_\ell^{\text{obs} ,ij}= C_\ell^{ij} + \frac{\delta_{ij}}{n(i)}.
\ee
Since the spectra $C_\ell^{ij}$ form a symmetric matrix with $N_\text{Bins} (N_\text{Bins}+1)/2$ independent terms, the covariance matrix for each $\ell$ is of dimension $[N_\text{Bins} (N_\text{Bins}+1)/2]^2$. We will also consider an approximate version of the covariance matrix in which cross-correlation between bins are neglected: then the covariance matrix is only of dimension $[N_\text{Bins}]^2$.

\begin{figure}[tbp]
\begin{center}
\includegraphics[width=.75\columnwidth]{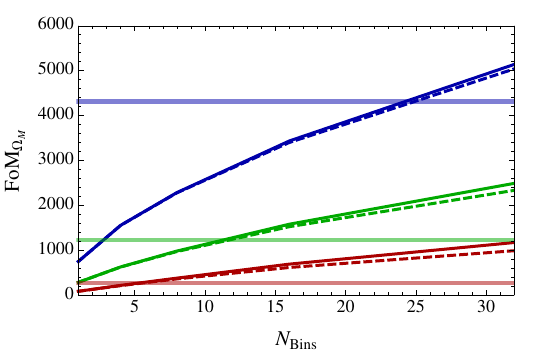}
\end{center}
\caption{FoM as a function of bin number, for a spectroscopic survey (like DESspec). The analysis is performed with a cut-off at different wavenumbers (from top to bottom): $k_\text{max} = 0.2 \ h \ \text{Mpc}^{-1}$ (blue),  $k_\text{max} = 0.1 \ h \ \text{Mpc}^{-1}$ (green),  $k_\text{max} = 0.05 \ h \ \text{Mpc}^{-1}$ (red). Different line styles show the figure of merit (FoM) considering all the cross correlation spectra between redshift bins (solid lines) or only auto-correlation spectra within redshift bins (dashed). The horizontal lines show the FoM computed for an analysis based on the 3D Fourier spectrum $P(k,\bar{z}_i)$, defined in detail in~\cite{Tegmark:1997rp,Didio2}. In the latter case the dependence on $N_\mathrm{Bins}$ is negligible.
}
\label{fig:FoM_OmegaCDM}
\end{figure}

In Fig.~\ref{fig:FoM_OmegaCDM} we show the Figure of Merit (FoM) for $\Omega_{m}$ when all other parameters are fixed, defined as the square root of the diagonal element of the Fisher matrix corresponding to $\Omega_m$ (see \cite{Didio2} for more details on this definition). We analyse how the FoM depends on the number of bins considering a spectroscopic survey (like DESspec) with a redshift range from $z=0.45$ to $z=0.65$. For comparison, we show the same FoM for an analysis of the observational data based on the reconstruction of the three-dimensional Fourier spectrum $P(k,\bar{z}_i)$ instead of the angular power spectra $C_\ell^{\text{obs} ,ij}$, using the same number of redshift bins in both cases~\cite{Tegmark:1997rp,Didio2}.

\begin{figure}[tbp]
\begin{center}
\includegraphics[width=.5\columnwidth]{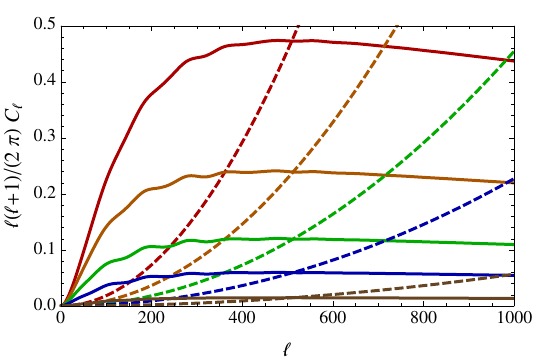}~
\includegraphics[width=.5\columnwidth]{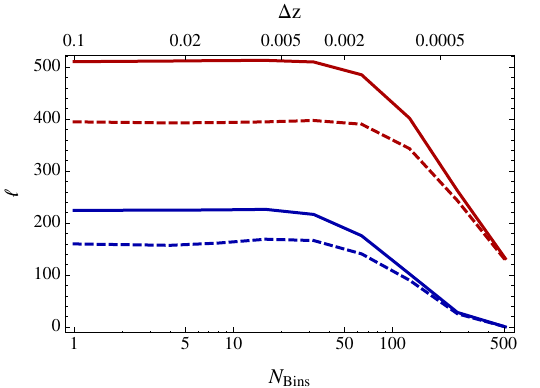}
\end{center}
\caption{In the left panel we show the angular power spectrum $C_\ell$ (solid lines) and the shot-noise contribution (dashed lines) for different top-hat window functions of half-widths: $\Delta z=0.1$ (brown), $\Delta z=0.025$ (blue), $\Delta z = 0.0125$ (green), $\Delta z = 0.00625$ (orange), $\Delta z = 0.003125$ (red). All the window functions are centered at $z=0.55$ and we consider a spectroscopic survey like DESspec for the shot-noise contribution (assuming the lowest shot noise mentioned in Footnote~\ref{footnoteSN}).
\newline
In the right panel we plot the multipole $\ell$ at which the shot-noise term starts to dominate. In red (upper lines) we consider low shot noise, while in blue (lower lines) higher shot noise is assumed (see Footnote~\ref{footnoteSN}). We consider top-hat (solid lines) and Gaussian window functions (dashed lines).}
\label{fig:winsize_vs_noise}
\end{figure}

Here we consider the same redshift binning strategy and the same shot-noise terms as in Ref.~\cite{Asorey+12}\footnote{\label{footnoteSN} The shot noise is determined by the galaxy number density, assumed to be $n=3.14 \times 10^{-3} h^3 \text{Mpc}^{-3}$. In Fig.~\ref{fig:winsize_vs_noise}, we also show for comparison some results, based on a larger shot noise assumption, with $n= 6.89 \times 10^{-4}h^3\text{Mpc}^{-3}$.}. We also consider the same set of non-linearity wavenumbers, $k_{\max}=  (0.05, 0.1, 0.2) h$Mpc$^{-1}$. We limit $\ell <\ell_{\max}$ such that scales orthogonal to the line-of-sight and smaller than $\la_{\min}=2\pi/k_{\max}$ (on which non-linearities are important) are not considered. 
This condition is equivalent to $[{2 \pi}/{\ell_{\max}}] D_A(\bar{z}) = a(\bar{z}) [{2\pi}/{k_{\max}}]$, giving $\ell_{\max} = r (\bar z) k_{\max}$ in flat space.
As shown in Fig.~\ref{fig:FoM_OmegaCDM}, more redshift bins lead to better constraints on the parameter $\Omega_m$. Instead, the FoM of the 3D Fourier spectrum analysis is almost independent of $N_\mathrm{Bins}$. The approach based on the angular power spectrum performs significantly better beyond a certain value of $N_\mathrm{Bins}$ depending on $k_{\max}$, that can be read off from the figure.
Our results cannot be immediately compared to those of~\cite{Asorey+12} due to a different normalization of the FoM (see~\cite{Didio2}). However they are consistent with~\cite{Asorey+12}, since the FoM of the two methods intersect each other at roughly the same value of $N_\mathrm{Bins}$.

Our results do not show any saturation when considering more and more redshift bins. Naively one might think that, by considering smaller redshift bins, the shot-noise term would start  to dominate at larger scales, since there are less galaxies per bin. But, as shown in Fig.~\ref{fig:winsize_vs_noise}, this effect is partially compensated by the growth of the signal which is integrated over a narrower window function and therefore less ``washed out''. Hence the scale at which the shot-noise term starts dominating changes very slowly when decreasing the width of the window function. 

According to Fig.~\ref{fig:winsize_vs_noise}, we expect to start seeing a saturation of the FoM between $50$ and $100$ redshift bins, which is above the range of values of $N_\mathrm{Bins}$ shown in  Fig.~\ref{fig:FoM_OmegaCDM}.
This comment applies to the noise of the autocorrelation spectrum in each bin. The shot noise of cross-correlations spectra in pairs of bins is actually twice smaller, so we expect the saturation to occur at an even larger number of bins.  
We have not computed the FoM up to this level, because it would be computationally very expensive (the calculation time of the FoM scales like $N_\text{Bins}^4$). Also, such a large number of bins probably becomes unrealistic as soon as one includes instrumental noise in the analysis.

\begin{figure}[tbp]
\begin{center}
\includegraphics[width=.5\columnwidth]{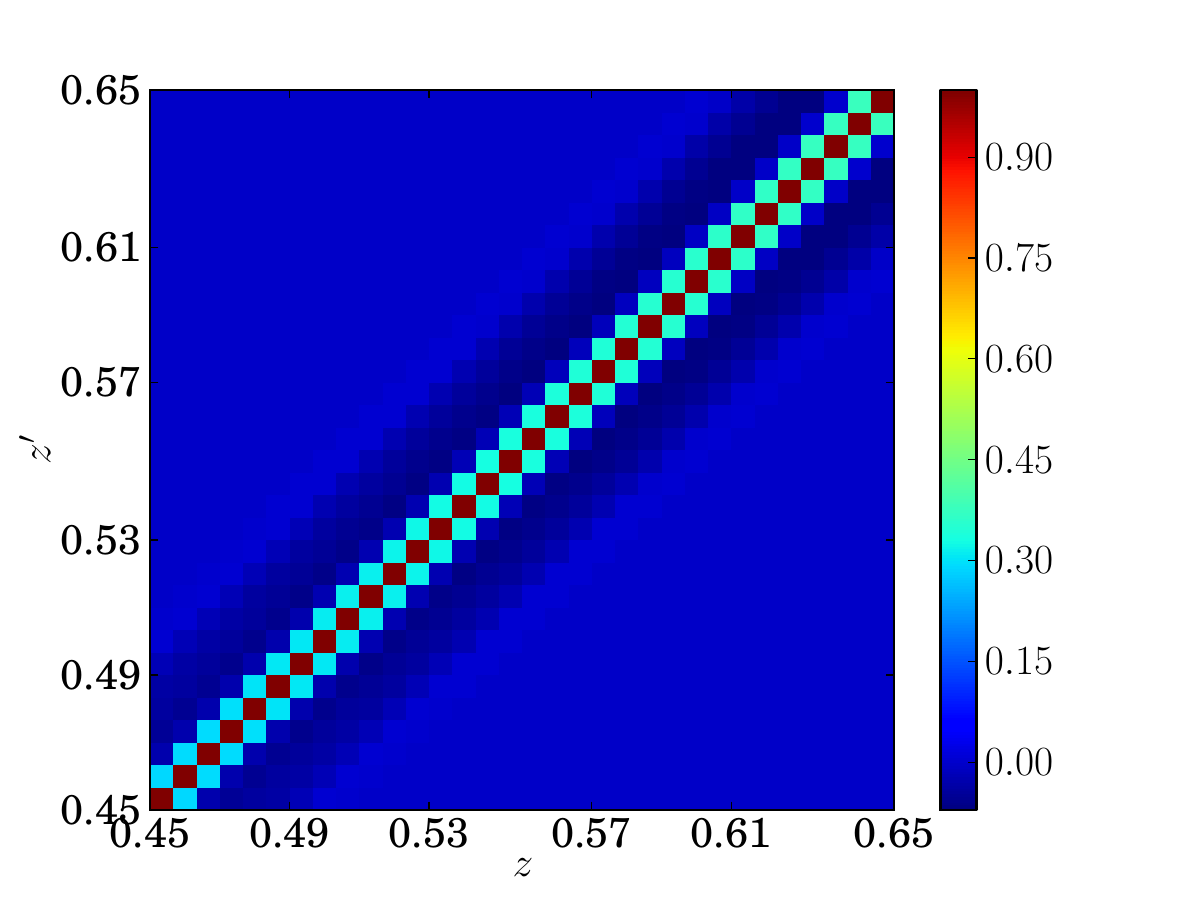}~
\includegraphics[width=.5\columnwidth]{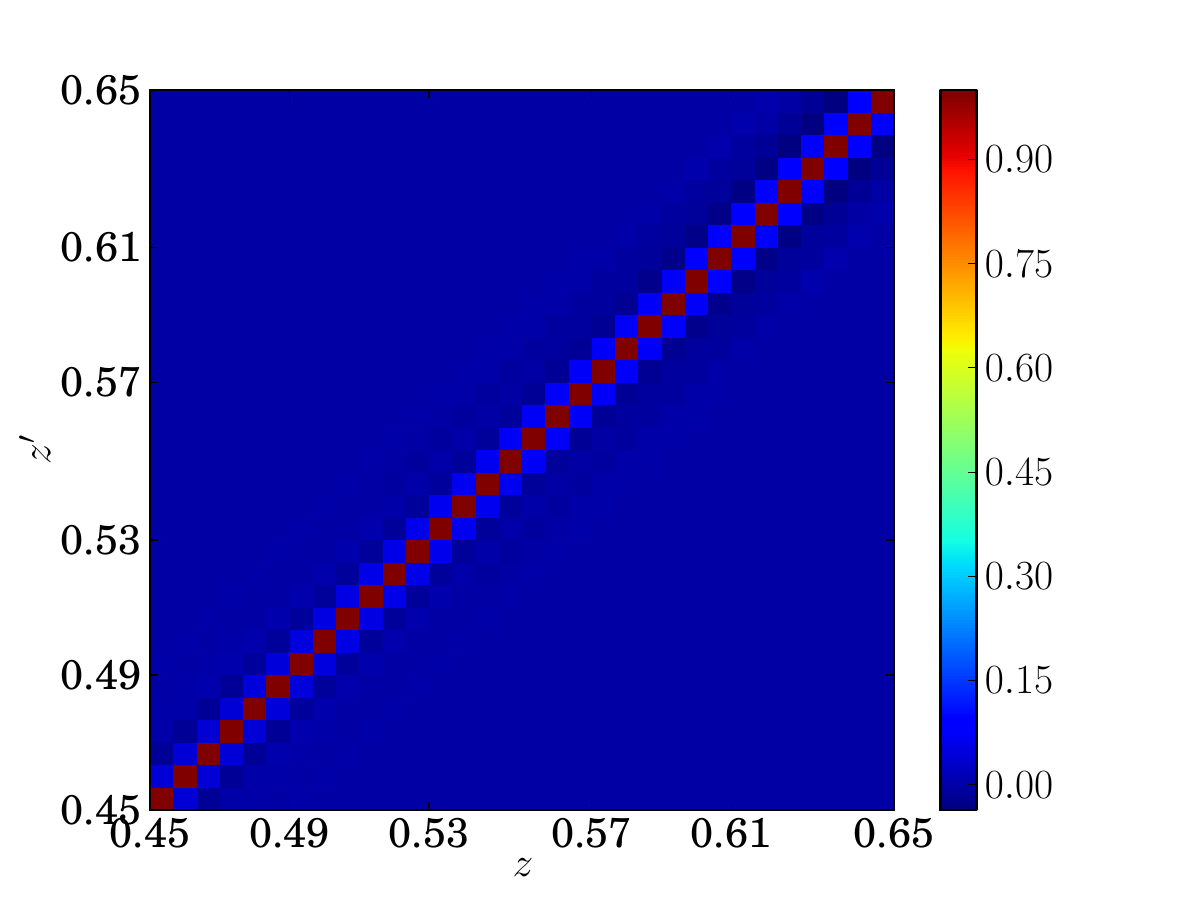}
\end{center}
\caption{We show the correlations between different redshift bins. The left panel is computed for $\ell=71$ corresponding to $k= 0.05 \ h \ \text{Mpc}^{-1}$, while the right panel shows $\ell=285$ corresponding to $k= 0.2 \ h \ \text{Mpc}^{-1}$ for the mean redshift $\bar z =0.55$. We consider a tophat window function with half-width $\Delta z = 0.003125$.
}
\label{fig:matrix_correlations}
\end{figure}

Finally, in Fig.~\ref{fig:matrix_correlations} we  show the covariance matrix at two different $\ell$ values. The largest signal in the correlation matrix comes clearly from the auto-correlation in each redshift bin.

\section{Conclusions and outlook}\label{s:con}

In this paper, we introduce a new version of the \class{} code which computes the linear angular power spectra for large scale structure, $C_\ell^{ij}$, in redshift bins described by a set of window functions $W_i(z)$. This code is called \classgal{} and includes all relativistic effects to first order in perturbation theory. The parts that concern the CMB have not been changed with respect to the main \class{} distribution. Several new features of \classgal{} will be merged with the main code in future \class{} versions.

The accuracy of the calculation is similar to the one of the original \class{} code, i.e., overall about 0.1\% when using default precision, or up to 0.01\% with boosted accuracy settings~\cite{Lesgourgues:2011rg}.
Whenever possible, we checked that \classgal{} agrees well with {\sc camb}sources. \classgal{} offers the advantage of including by default several window functions and the possibility of passing tabulated selection and evolution redshift distributions, remaining accurate and efficient for narrow redshift bins and large values of  $\ell$.
The code uses the Limber approximation as described in the Appendix, but for very accurate calculation one can turn off this approximation. For \class{} users, it should be absolutely straight forward to work with this code after reading the explanatory material presented in Appendix~\ref{app:A} of this work. A newcomer may want to read first the description of the original code in Ref.~\cite{Lesgourgues:2011re}.

We have illustrated the utility of this code with an example where we determine 
$\Om_m$ from a DES-like galaxy redshift survey. We have studied how the figure of merit depends on the number of redshift bins used, and to which extent slim redshift bins compensate for the increased shot noise by an enhanced signal. Our findings agree well with previous results~\cite{Asorey+12} based on {\sc camb}sources. More applications of our code are found in the accompanying paper~\cite{Didio2}.

\subparagraph{Acknowledgments:}
\hspace{0.5cm} We  are grateful to  Adam~Amara, Camille~Bonvin, Stefano~Camera, Enrique Gazta\~naga, Antony Lewis, Roy~Maartens, Andrina Nicola, Hideki~Perrier, Alexandre~Refregier and Carlo~Schimd, for helpful discussions and for comparison of codes. 
We acknowledge financial support by the Swiss National Science Foundation.
RD was supported in part by the (US) National Science Foundation under Grant No. NSF PHY11-25915.

\appendix 

\section{Differences between \class{} and \classgal}\label{app:A}

\subsection{Conventions and notations used in the code\label{conventions}}

The \class{} code uses the notation of Ma \& Bertschinger~\cite{Ma:1995ey} for both Newtonian and synchronous gauge (as usual, the latter is fully specified by requiring in addition that $\theta_\mathrm{CDM}$ vanishes at initial time, and hence at all times). Conformal time is denoted $\tau$ and the prime stands for $'\equiv \partial_\tau$. Instead of ${\cal H}$, the code uses the standard Hubble parameter $H=a'/a^2$ and its derivative $H'=a''/a^2- 2 a H^2$. Metric perturbations read $(\phi,\psi)$ in Newtonian gauge, and $(\eta, h)$ in synchronous gauge. The gauge-invariant Bardeen potentials  $(\Phi, \Psi)$ appearing in Eq.~(\ref{Dez}) are defined in such a way that in Newtonian gauge, they reduce to $(\phi,\psi)$, while in  synchronous gauge, they are given by $([\eta - {\cal H} \alpha], [{\cal H} \alpha + \alpha'])$, with $\alpha \equiv (h'+6\eta')/(2k^2)$~\cite{Ma:1995ey}. 

The quantities integrated by the code are not actual Fourier modes depending on $\vec{k}$, but transfer functions depending on $k$, normalized with respect to the curvature perturbation ${\cal R}$ (to be precise, this is true in the case of adiabatic initial conditions; for isocurvature initial conditions \class{} uses standard normalization conventions that can easily be read from the code). Hence, for any perturbation $A(\tau,\vec{k})$ with adiabatic initial conditions, the transfer function $A(\tau,k)$ is defined as
\begin{equation}
A(\tau,k) \equiv \frac{A(\tau,\vec{k})}{{\cal R}(\tau_\mathrm{ini},\vec{k})}~.
\end{equation}
The power spectrum of $A$ is then related to the primordial curvature power spectrum by
\begin{equation}
\langle A(\tau,\vec{k}) A^*(\tau,\vec{k}') \rangle = A(\tau,k)^2 P_{\cal R}(k) \delta^{(3)}(\vec{k}-\vec{k}')~,
\end{equation}
and the dimensionless primordial power spectrum is defined as ${\cal P}_{\cal R}(k) = \frac{k^3}{2\pi^2} P_{\cal R}(k)$.

\subsection{Modifications to the {\tt input} module}
\label{s:input}
\classgal{} incorporates a few more optional input parameters than \class. In order to compute the angular power spectra including relativistic corrections, one should include at least {\tt output = rCl, ...} in the input parameter file, instead of the usual flag {\tt dCl} referring to the same quantity without relativistic corrections computed also by the main \class. The three fields {\tt rCl\_rsd}, 
{\tt rCl\_lensing}, {\tt rCl\_gr} are set by default to {\tt yes}, but by setting one or several of them to {\tt no}, one can turn off the contribution of redshift-space distortions + Doppler, of lensing, or of gravitational potential terms.
The user can  pass a value for linear, scale-independent galaxy bias (e.g. {\tt bias = 1.2}). The density transfer function defined below in the first line of Eq.~(\ref{eq:delta_terms}) is multiplied everywhere by this factor.
Magnification bias, by default set to zero, is given by the option {\tt s\_bias} that allows constant values.
Source evolution is controlled by the field {\tt dNdz\_evolution} (by default blank and hence neglecting evolution), that represents the number of sources per redshift and solid angle as a function of redshift.
A tabulated evolution function can be considered by passing its file name, e.g., {\tt dNdz\_evolution = myevolution.dat}.

The minimum and maximum value of $\ell$ that will be computed are set by {\tt l\_max\_lss} and {\tt l\_min}. The latter is set by default to 2 but can be decreased to 1 or 0. For $\ell=0$ the usual output $[\ell(\ell+1)/2\pi] C_\ell$ would of course vanish, so there is an option {\tt cl\_rescale = yes/no} : if this is set to {\tt no}, the code will simply output the $C_\ell$'s.

The shape and characteristics of window functions can be set like in the main \class, i.e. by a sequence of the type:\\
{\tt
\mbox{ }selection = gaussian\\
\mbox{ }selection\_mean = 1, 1.5, 1.8\\
\mbox{ }selection\_width = 0.5,0.4,0.2\\
}
where the shape can be set to {\tt gaussian}, {\tt tophat} or {\tt dirac} (see the comments in the file {\tt explanatory.ini} for more details). These window functions can be multiplied by a selection function common to all redshift bins. This feature is disabled when the field {\tt dNdz\_selection} is left blank. For an analytic selection function hard-coded in the source files, one should write {\tt dNdz\_selection = analytic}: by default it will point to the function $dN/dz$ used in this paper. For reading a tabulated selection function from a file, one should pass the file name, e.g. {\tt dNdz\_selection = myselection.dat}. 

The user is free to use the same tabulated redshift distribution for the evolution and the selection function, i.e. the survey observes all the sources. One should simply use the same file for {\tt dNdz\_evolution} and {\tt dNdz\_selection}.

In the matrix  $C_{\ell}^{ij}$, the user may wish to neglect some non-diagonal terms $i\neq j$ in order to speed up the code and get more compact output files. The number of non-diagonal elements is set by {\tt non\_diagonal}, that can be assigned between 0 and $N-1$, where $N$ is the number of bins: 0 means `only auto-correlations', 1 means `only auto-correlations and adjacent bins', etc.

\subsection{Modifications to the {\tt perturbation} module}

The role of the perturbation module is to integrate the coupled system of evolution equations  for cosmological perturbations, and to store in memory a list of source functions $S_X(k,\tau)$ for discrete values of $k$ and $\tau$. These source functions are all linear combinations of transfer functions $A(\tau,k)$. In order to compute unlensed CMB spectra, one needs three well-known source functions $S_{T,E,B}$ described e.g. in \cite{Seljak:1996is,Tram:2013ima}.

The perturbation module can store many other source functions $S_X(k,\tau)$, depending on the requested output. In the main code \class{} v1.7, they consist in: 
\begin{itemize}
\item
individual density or velocity transfer functions  $\left\{ {\delta_i}(\tau,k),  {\theta_i}(\tau,k)\right\}$ (expressed in the gauge selected by the user); 
\item metric fluctuations if the lensing spectrum or lensed CMB spectrum are requested (the standard code stores only $S_g \equiv \Psi$ and assumes $\Phi=\Psi$, which is a good approximation for $\La$CDM at late times, but not sufficient for the purposes of this work); 
\item to prepare the computation of the Fourier matter power spectrum, $P(k)$, or of the harmonic power spectrum of matter density in shells, $C_\ell^{\delta_i \delta_j}$, the code can either store the total fluctuations of non-relativistic matter in a given gauge, $S_{\delta_m}={\delta_m}$, or the gravitational potential, $S_g = \Psi$, in view of inferring the total density fluctuation from the Poisson equation (in the sub-Hubble limit and assuming $\Phi=\Psi$). The user can use a flag to switch between these two schemes. Both of them provide approximations to the true observable density power spectrum built from Eq.~(\ref{Dez}), that we wish to compute with \classgal.
\end{itemize}

In \classgal, we need to store additional source functions corresponding to different terms in Eq.~(\ref{Dez}). These include the gauge-invariant matter density source function
(including CDM, baryons and non-relativistic neutrinos -- dubbed more generally non-cold dark matter ({\tt ncdm}) in the code),
\be
S_\mathrm{D} = \delta \rho_\mathrm{m} / \bar{\rho}_\mathrm{m} + 3 \frac{aH}{k^2} \theta_\mathrm{m}~,~~~~ {\rm (Newt.~or~synch.)}
\ee
the gauge-invariant velocity source function
\begin{eqnarray}
S_{\Theta} &=& \theta_\mathrm{m}  ~,~~~~~~~~~~~~~{\rm (Newt.)}\\
S_{\Theta} &=& \theta_\mathrm{m} + k^2 \alpha ~,~~~~{\rm (synch.)}
\end{eqnarray}
the gauge-invariant Bardeen potentials
\begin{eqnarray}
S_\Psi &=& \{ \psi ~~\mathrm{or}~~ {\cal H} \alpha + \alpha'\}~, \\
S_\Phi &=& \{\phi ~~\mathrm{or}~~ \eta - {\cal H} \alpha\}~,
\end{eqnarray}
and the sum $S_{(\Phi+\Psi)}=S_\Phi + S_\Psi$. The source functions $S_\mathrm{D}$ and $S_\Theta$ coincide with $D$ and $kV$ in Eq.~(\ref{Fl}). 
The code also needs to store the time derivatives  $S_{\Phi'} $ and $S_{(\Phi+\Psi)'}$.
In order to avoid heavy equations, these are not inferred from complicated differential combinations of the Einstein equations, but  from finite differences: $S_{\Phi'}(\tau)  \simeq [S_{\Phi'}(\tau+d \tau) - S_{\Phi'}(\tau-d \tau)]/[2 d \tau]$. 
Since both $\Phi$ and $\Psi$ vary only very slowly in the matter and dark energy dominated eras, this does not compromise the accuracy of the code.

\subsection{Modifications to the {\tt transfer} module}

The role of the transfer module is to calculate harmonic transfer functions $\Delta_\ell^X(k)$ by convolving the source functions $S_X(\tau,k)$ with Bessel functions, sometimes using a 
kernel (accounting for selection functions, rescaling factors, etc.) For instance, in the standard version of \class{} and in the flat space limit, the density transfer function $\Delta_{\ell}^{\delta_i}(k)$ in a given redshift bin is computed using the Poisson equation
\begin{equation}
-\frac{k^2}{a^2} \psi = 4 \pi G \rho_\mathrm{m} \delta_\mathrm{m} = \frac{3}{2} H^2 \Omega_\mathrm{m}(\tau) \delta_\mathrm{m} \;,
\end{equation}
Where $\Omega_\mathrm{m}(\tau)$ is the fractional density of matter at time $\tau$, and $\delta_\mathrm{m}$ denotes the matter density fluctuation in synchronous gauge (\class{} also allows computations in Newtonian gauge). This relation neglects pressure from massive neutrinos which is however very small in the redshift range $z < 3$ where we use it.
Then
\begin{eqnarray}
\Delta_{\ell}^{\delta_i}(k) &=&
\int d\tau \, W_i(\tau) \, \delta_\mathrm{m}(\tau,k) \, j_\ell\left(k(\tau_0-\tau)\right) \label{transfer_integral} \nonumber \\
&=& \int d\tau \, W_i(\tau) \, \frac{2 k^2}{3 (\Omega_\mathrm{m} a^2H^2)_\tau} \, S_\mathrm{g}(\tau,k)\,  j_\ell\left(k(\tau_0-\tau)\right)~. 
\end{eqnarray}
Here $W_i(\tau)$ stands for the selection function of the $i$th redshift bin of a given experiment, specified by the user. The selection functions readily available in \class{} are Gaussians, top-hat distributions or Dirac distributions in redshift space,  always normalized to $\int_0^\infty W_i(z)=1$. Since they represent a number of galaxies per redshift interval, $W_i(z)=dN/dz$, the associated function $W_i(\tau)$ is calculated by the code according to 
\begin{equation}
W_i(\tau)\equiv - \frac{dz}{d\tau} \, W_i(z) = H(\tau(z)) \, W_i(z)~.
\end{equation}  
Each selection function is associated with a mean redshift $\bar{z}_i\equiv \int_0^\infty dz \, z W_i(z)$, and with a characteristic conformal time $\bar{\tau}_i\equiv\tau(\bar{z}_i)$. The calculation can be sped up by using the Limber approximation for small angular scales (large $\ell$'s) and nearby shells (small $\bar{z}_i$).
We emphasize, however, that the Limber approximation must be used with care, since it may introduce significant errors especially for narrow $z$-window functions and for $z$-bin cross-correlations. 
The user can specify a value for the precision parameter $(\ell/z)_\mathrm{Limber}$ (set to 30 by default). 
When for given values of $i$ and $\ell$ the condition 
\begin{equation}
\ell > (\ell/z)_\mathrm{Limber} \,\, \bar{z}_i \label{Limber}
\end{equation} 
is satisfied, the code switches to
\begin{equation}
\Delta_{\ell}^{\delta_i}(k) = W_i(\tau_L) \frac{2 k}{3 (\Omega_\mathrm{m} a^2 H^2)_{\tau_L}} S_\mathrm{g}(\tau_L,k) 
\sqrt{\frac{\pi}{(2\ell+1)}}
\end{equation}
with $\tau_L\equiv \tau_0-\frac{\ell+1/2}{k}$. This approximation corresponds to the first-order Limber approximation. The transfer module contains a routine allowing to switch to the second-order Limber approximation~\cite{LoVerde:2008re}, but we checked that the difference between the two is small when the condition (\ref{Limber}) is satisfied. The Limber approximation remains automatically switched off in the case of  Dirac selection functions, for which the integral of Eq.~(\ref{transfer_integral}) is replaced by
\begin{equation}
\Delta_{\ell}^{\delta_i}(k) =  \frac{2 k^2}{3 (\Omega_\mathrm{m} a^2H^2)_{\bar{\tau}_i}} \, S_\mathrm{g}(\bar{\tau}_i,k)\,  j_\ell\left(k(\tau_0-\bar{\tau}_i)\right)~.
\end{equation}
For high-precision calculations, the user can still avoid the Limber approximation for whatever selection function by setting $(\ell/z)_\mathrm{Limber}$ to a very large value.
For non-Dirac selection functions, the code automatically adapts the $\tau$-sampling of the source functions $S_X(k,\tau)$ to the width of the selection function, in order to perform an accurate integral in Eq.(\ref{transfer_integral}). Thin shells with a narrow $W_i(\tau)$ require a dense sampling of $S_X(k,\tau)$ in the vicinity of each $\bar{\tau}_i$, and increase the computation time and memory.
Still, the approach described in these paragraphs represents the fastest and simplest way to calculate approximate density power spectra in shells. However the goal of this paper and of \classgal{} is to use a more involved calculation, accounting for all relativistic correction.

For that purpose, we need to compute many other transfer functions in \classgal, corresponding to the different contributions in Eq.~(\ref{Fl}):
\begin{eqnarray}
\Delta_{\ell}^{\mathrm{Den}_i} &=& \int_0^{\tau_0} d\tau W_i \, b S_\mathrm{D} \, j_\ell \nonumber \\
\Delta_{\ell}^{\mathrm{Len}_i} &=& \ell(\ell+1) \int_0^{\tau_0} d\tau \, W^\mathrm{L}_i \,  S_{\Phi+\Psi} \, j_\ell \nonumber \\
\Delta_{\ell}^{\mathrm{D}1_i} &=& \int_0^{\tau_0} d\tau \, W_i \left( \frac{1 \! + \! \frac{H'}{aH^2} \! + \! \frac{2 -5s }{(\tau_0-\tau)aH  } +5s - f^N_{\rm evo}}{k}\right) S_{\Theta} \, \frac{d j_\ell}{dx}  \nonumber \\
\Delta_{\ell}^{\mathrm{D}2_i} &=& \int_0^{\tau_0} d\tau \, W_i \left(f^N_{\rm evo} -3\right)\frac{ aH}{k^2} S_{\Theta} \, j_\ell \nonumber \\
\Delta_{\ell}^{\mathrm{Red}_i} &=& \int_0^{\tau_0} d\tau \, W_i \left( \frac{1}{aH} \right) S_{\Theta} \, \frac{d^2j_\ell}{dx^2} \nonumber \\
\Delta_{\ell}^{\mathrm{G}1_i} &=& \int_0^{\tau_0} d\tau \, W_i \, \left(2 + \frac{H'}{aH^2}+\frac{2-5s}{(\tau_0-\tau)aH} +5s - f^N_{\rm evo} \right) S_\Psi \, j_\ell \nonumber \\
\Delta_{\ell}^{\mathrm{G}2_i} &=& \int_0^{\tau_0} d\tau \, W_i \left( -2 + 5s \right) S_\Phi \, j_\ell \nonumber \\
\Delta_{\ell}^{\mathrm{G}3_i} &=& \int_0^{\tau_0} d\tau \, W_i \, \left( \frac{1}{aH} \right) S_{\Phi'} \, j_\ell \nonumber \\
\Delta_{\ell}^{\mathrm{G}4_i} &=& \int_0^{\tau_0} d\tau \, W_i^{\mathrm{G}4} \,  S_{\Phi+\Psi} \, j_\ell \nonumber \\
\Delta_{\ell}^{\mathrm{G}5_i} &=& \int_0^{\tau_0} d\tau \, W_i^{\mathrm{G}5} \, S_{(\Phi+\Psi)'} \, j_\ell ~.
\label{eq:delta_terms}
\end{eqnarray}
We have omitted all the arguments: $k$ for the transfer functions, $(\tau,k)$ for the source functions, $x\equiv k(\tau_0-\tau)$ for the Bessel functions, and $\tau$ for selection and background functions. For the integrated terms `Len', G4 and G5, we have defined
\begin{eqnarray}
W_i^\mathrm{L}(\tau) &=& \int_0^\tau \!\! d\tilde{\tau} W_i(\tilde{\tau}) \left( \frac{2-5s}{2} \right) \frac{(\tau-\tilde{\tau})}{(\tau_0-\tau)(\tau_0-\tilde{\tau})} \nonumber \\
W_i^{\mathrm{G}4}(\tau) &=&  \int_0^\tau \!\! d\tilde{\tau} W_i(\tilde{\tau}) \frac{2 -5s}{(\tau_0-\tilde{\tau})} \\
W_i^{\mathrm{G}5}(\tau) &=& \int_0^\tau \!\! d\tilde{\tau} W_i(\tilde{\tau}) \left(1+\frac{H'}{aH^2} + \frac{2 -5s}{(\tau_0-\tilde{\tau}) aH} + 5s-f^N_{\rm evo} \right)_{\tilde{\tau}}~. \nonumber
\end{eqnarray}
These expressions are valid in flat space, and can be easily generalized to curved space by replacing the spherical Bessel functions by  hyper spherical Bessel functions (for an introduction, see~\cite{Durrer:2008aa}). 

The evolution term $f^N_{\rm evo}$ has been explicitly implemented in terms of the number of sources per redshift and solid angle $\bar n (z)$, i.e. {\tt dNdz\_evolution}, according to~\cite{Challinor:2011bk},
\be
f^N_{\rm evo} =  \frac{d}{\HH d \tau} \left( \ln \frac{\bar n(z) H}{\left( \tau_0 -\tau \right)^2}\right) =  \frac{H' }{aH^2}+ \frac{2 }{Ha \left( \tau_0-\tau\right)} -\frac{1}{a}  \frac{d \ln (\bar n (z))}{d z}.
\ee

The previous remarks concerning the Limber approximation, the Dirac selection function, and time sampling issues for other selection functions, apply equally to \class{} and \classgal.

\subsection{Modifications to the {\tt spectra} module}

The main task of the spectra module is to convolve the primordial spectrum with quadratic combinations of the transfer functions. For instance, in the standard \class{} version, auto-correlation ($i=j$) and cross-correlation ($i \neq j$) harmonic power spectra of matter density fluctuations in shells are given by
\begin{equation}
C_\ell^{\delta_i \delta_j} = 4 \pi
\int \frac{dk}{k}  {\cal P}_{\cal R}(k) \Delta_{\ell}^{\delta_i}(k) \Delta_{\ell}^{\delta_j}(k)~.
\end{equation}
In \classgal, we compute a similar expression, with $\Delta_{\ell}^{\delta_i}(k)$ replaced by the sum
\begin{eqnarray}
\Delta_{\ell}^{i}(k) 
&=& \Delta_{\ell}^{\mathrm{Den}_i}  ~~~~~~~~~~~~~\mathrm{(Density~term)} \nonumber \\
&+&  \Delta_{\ell}^{\mathrm{Len}_i}  ~~~~~~~~~~~~~\mathrm{(Lensing~term)} \nonumber \\
&+& \Delta_{\ell}^{\mathrm{D1}_i} + \Delta_{\ell}^{\mathrm{D2}_i} ~~~~\mathrm{(Doppler~term)} \\
&+& \Delta_{\ell}^{\mathrm{Red}_i}  ~~~~~~~~~~~~~\mathrm{(Redshift~space~dist.)} \nonumber \\
&+& \Delta_{\ell}^{\mathrm{G}1_i} ... + \Delta_{\ell}^{\mathrm{G}5_i}.~\mathrm{(Gravity~terms)} \nonumber
\end{eqnarray}
To show that this expression coincides with Eqs.~(\ref{Fl}, \ref{delta}, \ref{Cl2}), we replace everywhere $H$ by ${\cal H}$, $\tau$ by $r=\tau_0-\tau$ and $\tilde{\tau}$ by $\tilde{r}=\tau_0-\tilde{\tau}$. Doing this we obtain
\begin{eqnarray}
&&\hspace*{-2mm} \Delta_{\ell}^{i}(k) = \nonumber \\
&&\hspace*{2mm} \int_0^{\tau_0} \!\!\!\! dr  \left\{
W_i(r)  j_\ell(kr) \left[ b S_\mathrm{D}(r,k) + \left( \frac{{\cal H}'}{{\cal H}^2}+\frac{2 -5s}{r {\cal H}} +5s-f^N_{\rm evo} +1 \right)_r S_\Psi(r,k)  \right. \right. \nonumber \\
&& \hspace*{3.4cm}\left.+ \left( - 2 +5s \right)S_\Phi(r,k) + \frac{1}{{\cal H}(r)} S_{\Phi'}(r,k) \right]  \nonumber \\
&& \hspace*{1.1cm}+~W_i(r)  \left[ \left. \frac{dj_\ell}{dx}\right|_{kr} \left( \frac{{\cal H}'}{{\cal H}^2} + \frac{2 -5s}{r {\cal H}} +5s-f^N_{\rm evo} \right)_r
+ \left. \frac{d^2j_\ell}{dx^2}\right|_{kr} \frac{k}{{\cal H}(r)}  +(f^N_{\rm evo}-3)j_\ell(kr) \frac{\cal H}{k} \right]  \frac{S_\Theta(r,k)}{k} 
\nonumber \\
&&
 \hspace*{1.1cm}+~\int_r^{\tau_0} \!\!\!\! d \tilde{r}  \, W_i(\tilde{r}) \, j_\ell(kr) \, \left[ \left(\frac{2 -5s}{2} \right) \left(\ell(\ell+1) \frac{\tilde{r}-r}{\tilde{r} r} + \frac{2}{\tilde{r}} \right) S_{\Phi+\Psi}(r,k) \nonumber \right. \\
&& \left. \left. \hspace*{4.4cm}+~\left.\left(\frac{{\cal H}'}{{\cal H}^2} + \frac{2 -5s}{\tilde{r} {\cal H}} +5s-f^N_{\rm evo}\right)\right|_{\tilde{r}}S_{(\Phi+\Psi)'}(r,k) \right] 
\right\}     ~. \nonumber \\
\end{eqnarray}
We can invert the order of the integrals over $r$ and $\tilde{r}$, and rename the integration variables: we replace $r$ in the first two lines and $\tilde{r}$ in the last line with $r_S$.  With this, the expression becomes
\begin{eqnarray}
&&\Delta_{\ell}^{i}(k) = \nonumber \\
&&\int_0^{\tau_0} \!\!\!\! dr_S W_i(r_S) \left\{
j_\ell(kr_S) \left[ b S_\mathrm{D}(r_S,k) +  \left( \frac{{\cal H}'}{{\cal H}^2}+\frac{2 -5s}{r_S {\cal H}} +5s-f^N_{\rm evo} +1 \right)_{r_S} \!\!\!\!S_\Psi(r_S,k)  \right.\right. \nonumber \\
&&\left.\left.\hspace*{3.6cm} +~\left( -2 +5s \right)S_\Phi(r_S,k) + \frac{1}{{\cal H}(r_S)} S_{\Phi'}(r_S,k) \right]  \right. \nonumber \\
&& \hspace*{1.5cm}+~\left[ \left. \frac{dj_\ell}{dx}\right|_{kr_S} \left( \frac{{\cal H}'}{{\cal H}^2} + \frac{2 -5s}{r_S {\cal H}} +5s-f^N_{\rm evo} \right)_{r_S}
+ \left. \frac{d^2j_\ell}{dx^2}\right|_{kr_S} \frac{k}{{\cal H}(r_S)} \right. \nonumber \\
&&\left.\hspace*{2.cm} +(f^N_{\rm evo}-3)j_\ell(kr_S) \frac{\cal H}{k} \right]  \frac{S_\Theta(r_S,k)}{k} 
\nonumber \\
&& \hspace*{1.5cm}+~\int_0^{r_S} \! dr \, j_\ell(kr) \, \left[ S_{\Phi+\Psi}(r,k)  \left( \frac{2-5s}{2}\right)\left(\ell(\ell+1) \frac{r_S-r}{r_S r} + \frac{2}{r_S} \right)\nonumber \right. \\
&& \left. \left. \hspace*{4.1cm}+~S_{(\Phi+\Psi)'}(r,k) \left.\left(\frac{{\cal H}'}{{\cal H}^2} + \frac{2 -5s}{r_S {\cal H}} +5s- f^N_{\rm evo} \right)\right|_{r_S} 
\right] \right\}    ~. \nonumber \\
\end{eqnarray}
which is identical to the thin shell expression in Eq.~(\ref{Fl}), with an additional integration over the window function, $dr_S W_i(r_S)$.

\section{Luminosity fluctuations}
\label{Lum_fluct_Appendix}

In this appendix we derive in detail the luminosity fluctuation used in expression (\ref{DezF}). We use the fact that the fractional fluctuation in the luminosity at fixed flux is given by twice the fractional fluctuation in the luminosity distance, 
\be \nonumber
\frac{\delta L_S}{\bar L_S} = 2 \frac{\delta D_L}{\bar D_L}.
\ee
We start from the luminosity distance fluctuation derived in \cite{Bonvin:2005ps} and we change the integration variable from the conformal time $\tau$ to $r = \tau_0 - \tau$
\bea \label{dist_1}
\frac{\delta D_L}{\bar D_L} &=& \left( \frac{1}{r_S \HH_S} -1 \right) \left( {\bf v}_S \cdot {\bf n} + \Psi_S \right) + \frac{1}{2 r_S}\int_0^{r_S} dr \left[ 2 - \frac{\left( r_S - r \right)}{rr_S }\Delta_\Omega \right] \left( \Psi + \Phi \right) \nonumber \\
&&\hspace{-.8cm}+ \frac{1}{r_S \HH_S} \int_0^{r_S} dr \left( \Psi' + \Phi' \right) - \int_0^{r_S} dr \frac{r_S -r}{r_S} \left( \Psi' + \Phi' \right) + \frac{1}{2 r_S}\int_0^{r_S} dr \left( r_S -r \right) r \left(  \Psi'' + \Phi'' \right) \nonumber \\
&&\hspace{-.8cm}-~ \frac{1}{2 r_S}\int_0^{r_S} dr \left( r_S - r \right) r \left[ \partial_r^2 +~ \frac{2}{r}\partial_r \right] \left( \Psi + \Phi \right)  + \frac{\Psi_S - \Phi_S}{2}
\eea
where we have neglected the local monopole and dipole terms and we have written the Laplacian in spherical coordinates,
$\De = \dd^2_r +\frac{2}{r}\dd_r +\frac{1}{r^2}\De_\Om$.  We have also used ${\bf n } \cdot {\bf \nabla}=-\dd_r$. The last term is not present in \cite{Bonvin:2005ps}, since there it is assumed that $\Psi= \Phi$, however, it can be found e.g. in Ref.~\cite{EneaMaster}. Considering the total derivative along the geodesic path
\be
\frac{d \ f \! \left( \tau, {\bf x} \left( \tau \right) \right)}{dr} =-\frac{d \ f \! \left( \tau, {\bf x} \left( \tau \right) \right)}{d\tau}= - f' - {\bf n } \cdot {\bf \nabla} f = -f' + \partial_r f.
\ee
we can rewrite the last integral of (\ref{dist_1}) as
\bea
&&\hspace*{-1cm}\frac{1}{2 r_S}\int_0^{r_S} dr \left( r_S - r \right) r \left[ \partial_r^2 + \frac{2}{r}\partial_r \right] \left( \Psi + \Phi \right) \nonumber \\
&&= \frac{1}{2 r_S}\int_0^{r_S} dr \left( r_S - r \right) r \left[ \frac{d^2}{dr^2}  + 2 \frac{d \partial_\tau }{dr} + \partial^2_\tau + \frac{2}{r} \frac{d}{dr} + \frac{2}{r} \partial_\tau \right] \left( \Psi + \Phi \right) \nonumber \\
&&= \frac{ \Psi_S + \Phi_S}{2}+ \frac{1}{2 r_S} \int_0^{r_S} dr \left( r_S - r \right) r \left( \Psi'' + \Phi'' \right) + \int_0^{r_S} dr   \frac{r}{r_S}  \left( \Psi' + \Phi' \right) \,.
\eea
Combining all terms together we finally arrive at
\bea
\frac{\delta D_L}{\bar D_L} &=& \left( \frac{1}{r_S \HH_S} -1 \right) \left( {\bf v}_S \cdot {\bf n} + \Psi_S  + \int_0^{r_S} dr \left( \Psi' + \Phi' \right)\right) \nonumber \\
&+& \frac{1}{2 r_S}\int_0^{r_S} dr \left[ 2 - \frac{\left( r_S - r \right) }{r }\Delta_\Omega \right] \left( \Psi + \Phi \right)  - \Phi_S .\nonumber
\eea

\bibliography{CLASS-refs}

\providecommand{\href}[2]{#2}\begingroup\raggedright\begin{thebibliography}{10}

\bibitem{Komatsu:2010fb}
{\bf WMAP Collaboration} Collaboration, E.~Komatsu et~al., {\it {Seven-Year
  Wilkinson Microwave Anisotropy Probe (WMAP) Observations: Cosmological
  Interpretation}},  {\em Astrophys. J. Suppl.} {\bf 192} (2011) 18,
  [\href{http://xxx.lanl.gov/abs/1001.4538}{{\tt arXiv:1001.4538}}].

\bibitem{Larson:2010gs}
D.~Larson, J.~Dunkley, G.~Hinshaw, E.~Komatsu, M.~Nolta, et~al., {\it
  {Seven-Year Wilkinson Microwave Anisotropy Probe (WMAP) Observations: Power
  Spectra and WMAP-Derived Parameters}},  {\em Astrophys. J. Suppl.} {\bf 192}
  (2011) 16, [\href{http://xxx.lanl.gov/abs/1001.4635}{{\tt arXiv:1001.4635}}].

\bibitem{Durrer:2008aa}
R.~Durrer, {\em The Cosmic Microwave Background}.
\newblock Cambridge University Press, {Cambridge, UK}, 2008.

\bibitem{Ade:2013ktc}
{\bf Planck Collaboration} Collaboration, P.~Ade et~al., {\it {Planck 2013
  results. I. Overview of products and scientific results}},
  \href{http://xxx.lanl.gov/abs/1303.5062}{{\tt arXiv:1303.5062}}.

\bibitem{Ade:2013zuv}
{\bf Planck Collaboration} Collaboration, P.~Ade et~al., {\it {Planck 2013
  results. XVI. Cosmological parameters}},
  \href{http://xxx.lanl.gov/abs/1303.5076}{{\tt arXiv:1303.5076}}.

\bibitem{BeltranJimenez:2010bb}
J.~Beltran~Jimenez and R.~Durrer, {\it {Effects of biasing on the matter power
  spectrum at large scales}},  {\em Phys.Rev.} {\bf D83} (2011) 103509,
  [\href{http://xxx.lanl.gov/abs/1006.2343}{{\tt arXiv:1006.2343}}].

\bibitem{Lesgourgues:2011re}
J.~Lesgourgues, {\it {The Cosmic Linear Anisotropy Solving System (CLASS) I:
  Overview}},  \href{http://xxx.lanl.gov/abs/1104.2932}{{\tt arXiv:1104.2932}}.

\bibitem{Blas:2011rf}
D.~Blas, J.~Lesgourgues, and T.~Tram, {\it {The Cosmic Linear Anisotropy
  Solving System (CLASS) II: Approximation schemes}},  {\em JCAP} {\bf 1107}
  (2011) 034, [\href{http://xxx.lanl.gov/abs/1104.2933}{{\tt
  arXiv:1104.2933}}].

\bibitem{Yoo:2009au}
J.~Yoo, A.~L. Fitzpatrick, and M.~Zaldarriaga, {\it {A New Perspective on
  Galaxy Clustering as a Cosmological Probe: General Relativistic Effects}},
  {\em Phys.Rev.} {\bf D80} (2009) 083514,
  [\href{http://xxx.lanl.gov/abs/0907.0707}{{\tt arXiv:0907.0707}}].

\bibitem{Yoo:2010ni}
J.~Yoo, {\it {General Relativistic Description of the Observed Galaxy Power
  Spectrum: Do We Understand What We Measure?}},  {\em Phys.Rev.} {\bf D82}
  (2010) 083508, [\href{http://xxx.lanl.gov/abs/1009.3021}{{\tt
  arXiv:1009.3021}}].

\bibitem{Bonvin:2011bg}
C.~Bonvin and R.~Durrer, {\it {What galaxy surveys really measure}},  {\em
  Phys.Rev.} {\bf D84} (2011) 063505,
  [\href{http://xxx.lanl.gov/abs/1105.5280}{{\tt arXiv:1105.5280}}].

\bibitem{Challinor:2011bk}
A.~Challinor and A.~Lewis, {\it {The linear power spectrum of observed source
  number counts}},  {\em Phys.Rev.} {\bf D84} (2011) 043516,
  [\href{http://xxx.lanl.gov/abs/1105.5292}{{\tt arXiv:1105.5292}}].

\bibitem{Bertacca:2012tp}
D.~Bertacca, R.~Maartens, A.~Raccanelli, and C.~Clarkson, {\it {Beyond the
  plane-parallel and Newtonian approach: Wide-angle redshift distortions and
  convergence in general relativity}},  {\em JCAP} {\bf 1210} (2012) 025,
  [\href{http://xxx.lanl.gov/abs/1205.5221}{{\tt arXiv:1205.5221}}].

\bibitem{Didio2}
E.~Di~Dio, F.~Montanari, R.~Durrer, and J.~Lesgourgues, {\it {Cosmological
  Parameter Estimation with Large Scale Structure Observations}},
  \href{http://xxx.lanl.gov/abs/1308.6186}{{\tt arXiv:1308.6186}}.

\bibitem{Kaiser1987}
N.~{Kaiser}, {\it {Clustering in real space and in redshift space}},  {\em
  M.N.R.A.S.} {\bf 227} (July, 1987) 1--21.

\bibitem{Hamilton1992}
A.~J.~S. {Hamilton}, {\it {Measuring Omega and the real correlation function
  from the redshift correlation function}},  {\em Astrophys. J.} {\bf 385}
  (Jan., 1992) L5--L8.

\bibitem{Raccanelli:2010hk}
A.~{Raccanelli}, L.~{Samushia}, and W.~J. {Percival}, {\it {Simulating
  redshift-space distortions for galaxy pairs with wide angular separation}},
  {\em M.N.R.A.S.} {\bf 409} (Dec., 2010) 1525--1533,
  [\href{http://xxx.lanl.gov/abs/1006.1652}{{\tt arXiv:1006.1652}}].

\bibitem{Montanari:2012me}
F.~Montanari and R.~Durrer, {\it {A new method for the Alcock-Paczynski test}},
   {\em Phys.Rev.} {\bf D86} (2012) 063503,
  [\href{http://xxx.lanl.gov/abs/1206.3545}{{\tt arXiv:1206.3545}}].

\bibitem{Lima:2010tq}
M.~Lima, B.~Jain, M.~Devlin, and J.~Aguirre, {\it {Submillimeter Galaxy Number
  Counts and Magnification by Galaxy Clusters}},  {\em Astrophys.J.} {\bf 717}
  (2010) L31, [\href{http://xxx.lanl.gov/abs/1004.4889}{{\tt
  arXiv:1004.4889}}].

\bibitem{Hezaveh:2010zk}
Y.~D. Hezaveh and G.~P. Holder, {\it {Effects of Strong Gravitational Lensing
  on Millimeter-Wave Galaxy Number Counts}},  {\em Astrophys.J.} {\bf 734}
  (2011) 52, [\href{http://xxx.lanl.gov/abs/1010.0998}{{\tt arXiv:1010.0998}}].

\bibitem{Zucca:2006}
E.~{Zucca}, O.~{Ilbert}, S.~{Bardelli}, L.~{Tresse}, G.~{Zamorani},
  S.~{Arnouts}, L.~{Pozzetti}, M.~{Bolzonella}, H.~J. {McCracken},
  D.~{Bottini}, B.~{Garilli}, V.~{Le Brun}, O.~{Le F{\`e}vre}, D.~{Maccagni},
  J.~P. {Picat}, R.~{Scaramella}, M.~{Scodeggio}, G.~{Vettolani},
  A.~{Zanichelli}, C.~{Adami}, M.~{Arnaboldi}, A.~{Cappi}, S.~{Charlot},
  P.~{Ciliegi}, T.~{Contini}, S.~{Foucaud}, P.~{Franzetti}, I.~{Gavignaud},
  L.~{Guzzo}, A.~{Iovino}, B.~{Marano}, C.~{Marinoni}, A.~{Mazure},
  B.~{Meneux}, R.~{Merighi}, S.~{Paltani}, R.~{Pell{\`o}}, A.~{Pollo},
  M.~{Radovich}, M.~{Bondi}, A.~{Bongiorno}, G.~{Busarello}, O.~{Cucciati},
  L.~{Gregorini}, F.~{Lamareille}, G.~{Mathez}, Y.~{Mellier}, P.~{Merluzzi},
  V.~{Ripepi}, and D.~{Rizzo}, {\it {The VIMOS VLT Deep Survey. Evolution of
  the luminosity functions by galaxy type up to z = 1.5 from first epoch
  data}},  {\em \aap} {\bf 455} (Sept., 2006) 879--890,
  [\href{http://xxx.lanl.gov/abs/astro-ph/0506393}{{\tt astro-ph/0506393}}].

\bibitem{Peng:2010}
Y.-j. {Peng}, S.~J. {Lilly}, K.~{Kova{\v c}}, M.~{Bolzonella}, L.~{Pozzetti},
  A.~{Renzini}, G.~{Zamorani}, O.~{Ilbert}, C.~{Knobel}, A.~{Iovino},
  C.~{Maier}, O.~{Cucciati}, L.~{Tasca}, C.~M. {Carollo}, J.~{Silverman},
  P.~{Kampczyk}, L.~{de Ravel}, D.~{Sanders}, N.~{Scoville}, T.~{Contini},
  V.~{Mainieri}, M.~{Scodeggio}, J.-P. {Kneib}, O.~{Le F{\`e}vre},
  S.~{Bardelli}, A.~{Bongiorno}, K.~{Caputi}, G.~{Coppa}, S.~{de la Torre},
  P.~{Franzetti}, B.~{Garilli}, F.~{Lamareille}, J.-F. {Le Borgne}, V.~{Le
  Brun}, M.~{Mignoli}, E.~{Perez Montero}, R.~{Pello}, E.~{Ricciardelli},
  M.~{Tanaka}, L.~{Tresse}, D.~{Vergani}, N.~{Welikala}, E.~{Zucca},
  P.~{Oesch}, U.~{Abbas}, L.~{Barnes}, R.~{Bordoloi}, D.~{Bottini}, A.~{Cappi},
  P.~{Cassata}, A.~{Cimatti}, M.~{Fumana}, G.~{Hasinger}, A.~{Koekemoer},
  A.~{Leauthaud}, D.~{Maccagni}, C.~{Marinoni}, H.~{McCracken}, P.~{Memeo},
  B.~{Meneux}, P.~{Nair}, C.~{Porciani}, V.~{Presotto}, and R.~{Scaramella},
  {\it {Mass and Environment as Drivers of Galaxy Evolution in SDSS and zCOSMOS
  and the Origin of the Schechter Function}},  {\em \apj} {\bf 721} (Sept.,
  2010) 193--221, [\href{http://xxx.lanl.gov/abs/1003.4747}{{\tt
  arXiv:1003.4747}}].

\bibitem{Bonvin:2005ps}
C.~Bonvin, R.~Durrer, and M.~A. Gasparini, {\it {Fluctuations of the luminosity
  distance}},  {\em Phys.Rev.} {\bf D73} (2006) 023523,
  [\href{http://xxx.lanl.gov/abs/astro-ph/0511183}{{\tt astro-ph/0511183}}].

\bibitem{Baldauf:2011bh}
T.~Baldauf, U.~Seljak, L.~Senatore, and M.~Zaldarriaga, {\it {Galaxy Bias and
  non-Linear Structure Formation in General Relativity}},  {\em JCAP} {\bf
  1110} (2011) 031, [\href{http://xxx.lanl.gov/abs/1106.5507}{{\tt
  arXiv:1106.5507}}].

\bibitem{Asorey+12}
J.~{Asorey}, M.~{Crocce}, E.~{Gazta{\~n}aga}, and A.~{Lewis}, {\it {Recovering
  3D clustering information with angular correlations}},  {\em \mnras} {\bf
  427} (Dec., 2012) 1891--1902, [\href{http://xxx.lanl.gov/abs/1207.6487}{{\tt
  arXiv:1207.6487}}].

\bibitem{McDonald:2008sh}
P.~McDonald and U.~Seljak, {\it {How to measure redshift-space distortions
  without sample variance}},  {\em JCAP} {\bf 0910} (2009) 007,
  [\href{http://xxx.lanl.gov/abs/0810.0323}{{\tt arXiv:0810.0323}}].

\bibitem{Bonvin2013}
C.~{Bonvin}, L.~{Hui}, and E.~{Gaztanaga}, {\it {Asymmetric galaxy correlation
  functions}},  {\em ArXiv e-prints} (Sept., 2013)
  [\href{http://xxx.lanl.gov/abs/1309.1321}{{\tt arXiv:1309.1321}}].

\bibitem{Crocce:2011}
M.~Crocce, A.~Cabr\'e, and E.~Gazta{\~n}aga, {\it {Modelling the angular
  correlation function and its full covariance in photometric galaxy surveys}},
   {\em M.N.R.A.S.} {\bf 414} (2011) 329--349,
  [\href{http://xxx.lanl.gov/abs/1004.4640}{{\tt arXiv:1004.4640}}].

\bibitem{Tegmark:1997rp}
M.~Tegmark, {\it {Measuring cosmological parameters with galaxy surveys}},
  {\em Phys.Rev.Lett.} {\bf 79} (1997) 3806--3809,
  [\href{http://xxx.lanl.gov/abs/astro-ph/9706198}{{\tt astro-ph/9706198}}].

\bibitem{Lesgourgues:2011rg}
J.~Lesgourgues, {\it {The Cosmic Linear Anisotropy Solving System (CLASS) III:
  Comparision with CAMB for LambdaCDM}},
  \href{http://xxx.lanl.gov/abs/1104.2934}{{\tt arXiv:1104.2934}}.

\bibitem{Ma:1995ey}
C.-P. Ma and E.~Bertschinger, {\it {Cosmological perturbation theory in the
  synchronous and conformal Newtonian gauges}},  {\em Astrophys.J.} {\bf 455}
  (1995) 7--25, [\href{http://xxx.lanl.gov/abs/astro-ph/9506072}{{\tt
  astro-ph/9506072}}].

\bibitem{Seljak:1996is}
U.~Seljak and M.~Zaldarriaga, {\it {A Line of sight integration approach to
  cosmic microwave background anisotropies}},  {\em Astrophys.J.} {\bf 469}
  (1996) 437--444, [\href{http://xxx.lanl.gov/abs/astro-ph/9603033}{{\tt
  astro-ph/9603033}}].

\bibitem{Tram:2013ima}
T.~Tram and J.~Lesgourgues, {\it {Optimal polarisation equations in FLRW
  universes}},  \href{http://xxx.lanl.gov/abs/1305.3261}{{\tt
  arXiv:1305.3261}}.

\bibitem{LoVerde:2008re}
M.~LoVerde and N.~Afshordi, {\it {Extended Limber Approximation}},  {\em
  Phys.Rev.} {\bf D78} (2008) 123506,
  [\href{http://xxx.lanl.gov/abs/0809.5112}{{\tt arXiv:0809.5112}}].

\bibitem{EneaMaster}
E.~{Di Dio}, {\it Master thesis},  ETH Z\"urich (2010).

\end{thebibliography}\endgroup
\bibliographystyle{JHEP}

\end{document}